\documentclass[11pt]{article}

\usepackage[margin=1in]{geometry}
\usepackage{graphicx}
\usepackage{multirow}
\usepackage{amsmath,amssymb,amsfonts}
\usepackage{amsthm}
\usepackage{mathrsfs}
\usepackage{xcolor}
\usepackage{textcomp}
\usepackage{booktabs}
\usepackage{dsfont}
\usepackage[round,authoryear]{natbib}
\usepackage[hidelinks]{hyperref}

\providecommand{\doi}[1]{doi:\,\href{https://doi.org/#1}{#1}}

\newcommand{\eps}{\varepsilon}

\newcommand{\R}{\mathds{R}}

\title{Continuous-time state space analysis of $\delta^{18}$O, $\delta^{13}$C, and CO$_2$ in the Cenozoic Era}

\author{
Mikkel Bennedsen$^{1,2}$,\quad Eric Hillebrand$^{1,2,*}$,\quad Siem Jan Koopman$^{3}$,\\
Kathrine Larsen$^{1,2}$,\quad Rachel Lupien$^{4}$ \\[1.5ex]
{\small $^{1}$Department of Economics and Business Economics, Aarhus University, Denmark}\\
{\small $^{2}$Center for Research in Energy: Economics and Markets (CoRE), Aarhus University, Denmark}\\
{\small $^{3}$Department of Econometrics, Vrije Universiteit Amsterdam, The Netherlands}\\
{\small $^{4}$Department of Geoscience, Aarhus University, Denmark}\\[1ex]
{\small $^{*}$Corresponding author: \texttt{ehillebrand@econ.au.dk}}
}

\date{\today}

\begin{document}

\maketitle

\begin{abstract}
\noindent We develop a continuous-time state-space framework for the joint reconstruction of three Cenozoic climate proxies, benthic foraminiferal $\delta^{18}$O and $\delta^{13}$C and atmospheric CO$_2$, from irregularly and unevenly sampled multi-site, multi-method data spanning the last 67 million years. The latent signals follow a trivariate random walk in continuous time; the measurement equation differentiates the error variance by drill site for the isotopes and by proxy group for CO$_2$, with bias intercepts placing all sources on a common scale, and the transition equation lets the innovation covariance and a deterministic La2004 Milankovitch forcing depend on the prevailing climate state. All parameters are estimated by maximum likelihood through the Kalman filter with diffuse initialization. The estimated cross-proxy correlations reverse sign between the early Cenozoic greenhouse and the icehouse, the orbital sensitivity of the isotopes strengthens as continental ice sheets grow, and the reconstructed CO$_2$ path, reported with calibrated confidence bands, places the atmospheric CO$_2$ thresholds of the major Cenozoic glaciations in relation to present-day concentrations.
\end{abstract}



\section{Introduction}\label{S:Intro}

Reconstructing the climate of the geological past means extracting latent signals from proxy records that are noisy, irregularly spaced in time, and assembled from many laboratories, sites, and measurement methods. These features are the natural province of state-space models and the Kalman filter, which separate signal from measurement error, accommodate missing and unevenly spaced observations, and deliver calibrated uncertainty for the reconstructed signal. \cite{bhkl2024} developed a continuous-time state-space model of this kind for benthic foraminiferal $\delta^{18}$O and $\delta^{13}$C over the Cenozoic. This paper extends that framework by adding atmospheric CO$_2$ as a third, jointly modeled proxy. The trivariate model retains the elements of the bivariate one (continuous-time random-walk dynamics, measurement variances and bias intercepts differentiated by source, a climate-state-dependent innovation covariance, and deterministic La2004 Milankovitch forcing of the transition equation with climate-state-dependent sensitivity), now estimated jointly across all three proxies.

The central contribution is the addition of CO$_2$ to the model. Extending the latent state from two dimensions to three is methodologically straightforward; the challenge lies instead in the structure of the CO$_2$ data. The CO$_2$ estimates are drawn from many proxy methods with systematic offsets and heterogeneous precision, they are observed at times that do not coincide with the isotope measurements, and they are far sparser than the isotopes (6{,}300 observations against roughly 24{,}000 of each isotope), with long gaps in the record. Jointly modeling proxies sampled at different times, at different densities, and with different error structures is therefore non-trivial. The continuous-time state-space formulation handles it naturally: each scalar observation is treated as a partial measurement of a shared latent vector, state innovations are added only over the time elapsed between successive observations (so that simultaneous and near-simultaneous measurements require no special treatment), and the measurement equation differentiates the error variance by drill site for the isotopes and by proxy group for CO$_2$, with site- and group-specific intercepts placing every source on a common latent scale. The sparsity of the CO$_2$ record lowers the precision of the CO$_2$-related estimates. The central idea of the joint approach is to mitigate this through the estimated cross-proxy correlations: the densely sampled $\delta^{18}$O and $\delta^{13}$C observations carry information about the latent CO$_2$ signal and can thereby help reconstruct it across the gaps in the CO$_2$ record.

The transition equation carries two forms of state dependence. The innovation covariance is differentiated across the six Cenozoic climate states of \cite{westerhold2020}, and the La2004 orbital solution \citep{laskar2004} enters as a deterministic forcing whose coefficients also vary across climate states, allowing the astronomical sensitivity of each proxy to strengthen as continental ice sheets develop. We use the estimated model to reconstruct Cenozoic CO$_2$ with calibrated confidence bands and to read off the atmospheric CO$_2$ levels associated with the three major glaciation events of the Cenozoic.

Many elements of the data sets we use are themselves estimates rather than direct measurements: the paleo-CO$_2$ values are reconstructed from geochemical proxies, and the ages are derived from astrochronological age models. Because we treat these compilations as the data on which our statistical models are estimated, we nonetheless refer to their individual elements as ``observations'' and to the compilations as ``data'' throughout the paper, and we reserve the language of estimation for the latent states and parameters of our models. This convention keeps it clear at every point whether we are speaking of the input data sets or of the model quantities estimated from them.

The remainder of the paper is organized as follows. Section~\ref{S:Data} describes the isotope and CO$_2$ data; Section~\ref{S:Model} sets out the univariate and trivariate models; Section~\ref{S:UniResults} reports the univariate CO$_2$ analysis; Section~\ref{S:TriResults} the trivariate estimates, including the orbital forcing; Section~\ref{S:Transitions} the reconstructed CO$_2$ across the major Cenozoic glaciations and the implied thresholds; and Section~\ref{S:Conclusion} concludes.


\section{Data}\label{S:Data}

\subsection{Benthic foraminifera isotope data}\label{S:DataIso}

We use the CENOGRID benthic foraminifera data set compiled by \cite{westerhold2020}, which provides measurements of $\delta^{18}$O and $\delta^{13}$C spanning the Cenozoic Era (the last 67 million years). The data set contains 24{,}321 time-stamped lines from 34 different studies, covering ages from 67.10 to 0.0006~Ma. After sorting by age, the data contain 24{,}259 observations of $\delta^{18}$O and 23{,}939 observations of $\delta^{13}$C, with the two proxies sharing the same set of time stamps. Of these 24{,}321 observations, only 23{,}722 fall on distinct time stamps; the difference of 599 arises because some time stamps carry more than one observation (up to four), on either or both series. For a detailed discussion of the isotope data, we refer to \cite{westerhold2020}.

\subsection{Paleo-CO$_2$ data}\label{S:DataCO2}

Reconstructions of atmospheric CO$_2$ over the Cenozoic rely on a variety of geochemical proxy methods, each with distinct assumptions, calibrations, and sources of uncertainty. We compile a merged data set from two releases of the Cenozoic CO$_2$ Proxy Integration Project \citep[CenCO2PIP;][]{cenco2pip2023}. The first component is the November~2023 ``product'' release \citep{cenco2pip_product2023}, which provides a vetted compilation of 5{,}698 paleo-CO$_2$ estimates from multiple proxy methods. The second component consists of 602~observations from the January~2026 data archive \citep{cenco2pip_archive2026} that are strictly new, i.e., they originate from studies not already represented in the November~2023 product. The union of these two sources gives a total of $N_{\text{CO}_2} = 6{,}300$ observations spanning 0 to 67~Ma. The product compilation spans all three CenCO2PIP confidence categories (estimates with fully developed uncertainty; estimates of high quality whose uncertainty is not yet fully constrained; and estimates from older methods or with incompletely quantified uncertainty).

The 6{,}300 observations originate from 141 individual studies and are classified into 13 proxy categories:
\begin{enumerate}
\item B/Ca ratios in foraminifera (\texttt{b\_ca}; 277 obs.),
\item boron isotopes (\texttt{boron\_isotopes}; 2{,}111 obs.),
\item land plant $\delta^{13}$C (\texttt{land\_plant\_d13c}; 1{,}799 obs.),
\item liverwort proxies (\texttt{liverwort}; 3 obs.),
\item nahcolite mineral proxies (\texttt{nahcolite}; 4 obs.),
\item paleosol carbonate proxies (\texttt{paleosol}; 319 obs.),
\item phytoplankton-based proxies (\texttt{phytoplankton}; 1{,}201 obs.),
\item stomatal proxies, Franks method (\texttt{stomata-franks}; 162 obs.),
\item stomatal proxies, Konrad FOM (\texttt{stomata-konrad-fom}; 30 obs.),
\item stomatal proxies, Konrad ROM (\texttt{stomata-konrad-rom}; 4 obs.),
\item stomatal proxies, SD (\texttt{stomata-sd}; 6 obs.),
\item stomatal index proxies (\texttt{stomata-si}; 294 obs.),
\item stomatal ratio proxies (\texttt{stomata-sr}; 90 obs.).
\end{enumerate}
The three largest categories (boron isotopes, land plant $\delta^{13}$C, and phytoplankton) together account for 81\% of all observations.

For the measurement-error model we pool these categories into seven groups: the six methods B/Ca, boron isotopes, land plant $\delta^{13}$C, nahcolite, paleosol, and phytoplankton are kept separate, and the six sparsely sampled stomatal sub-proxies together with the liverwort proxy are combined into a single stomata group. Pooling the sparse proxies keeps every group large enough to identify its own measurement variance and intercept.

CO$_2$ concentrations in the data set range from 45 to 3{,}560~ppm. There are 4{,}049 unique ages; the maximum number of observations at any single age is 258. The data are substantially denser in the more recent record: 2{,}963 observations fall in the Icehouse period (3.3--0~Ma) and 977 in Coolhouse~2 (13.9--3.3~Ma), whereas the three oldest climate states together account for only 1{,}306 observations.

We model the logarithm of CO$_2$ concentration, $y = \log(\text{CO}_2)$, for three reasons: (1) the distribution of CO$_2$ values is right-skewed with approximate log-normality, (2) relative (percentage) changes in CO$_2$ are more natural to compare across eras with very different baseline concentrations, and (3) the log transform stabilizes variance and ensures positivity of any reconstructed CO$_2$ path.


\section{Model}\label{S:Model}

\subsection{Univariate random walk plus noise for $\log(\text{CO}_2)$}\label{S:UniCO2}

We apply the continuous-time state space framework developed in \cite{bhkl2024} to the paleo-CO$_2$ record. In the simplest specification, the random-walk-plus-noise (RWN) model for $\log(\text{CO}_2)$ is
\begin{eqnarray}
y_n = \mu_{t_n} + \eps_n, & & \eps_n \stackrel{indep.}{\sim} \mathsf{N}(0, \sigma_{\eps}^2), \notag \\
\mu_{t_n + \Delta t_n} = \mu_{t_n} + \eta_{t_n}, & & \eta_{t_n} \stackrel{i.i.d.}{\sim} \mathsf{N}(0, \sigma_\eta^2 \Delta t_n), \label{E:RWN_CO2}
\end{eqnarray}
where $y_n = \log(\text{CO}_{2,n})$ is the observation at time $t_n$, $\mu_{t_n}$ is the latent signal, and $\Delta t_n = t_{n+1} - t_n \ge 0$ is the time increment. As in \cite{bhkl2024}, the variance of the state disturbance scales linearly with $\Delta t_n$, reflecting the Brownian motion dynamics of the random walk in continuous time.

The different proxy categories have distinct noise characteristics. Following the approach of differentiating measurement variances by drill site for the isotope data in \cite{bhkl2024}, we differentiate the measurement equation variance by proxy group for the CO$_2$ data, and add a group-specific intercept that places all groups on a common scale:
\begin{equation}
y_n = c_{g} + \mu_{t_n} + \eps_n, \qquad \eps_n \stackrel{indep.}{\sim} \mathsf{N}(0, \sigma_{\eps,g}^2), \qquad g = g(n) \in \{1, \ldots, 7\},
\label{E:RWN_CatV}
\end{equation}
where $g$ identifies the proxy group of observation $y_n$ and $c_g$ is its intercept (the bias correction), with $c_g = 0$ for the reference group \texttt{b\_ca}. Similarly, we differentiate the transition equation variance by climate state:
\begin{equation}
\mu_{t_n + \Delta t_n} = \mu_{t_n} + \eta_{t_n}, \qquad \eta_{t_n} \stackrel{i.i.d.}{\sim} \mathsf{N}(0, \sigma_{\eta,k}^2 \Delta t_n), \qquad k = k(n) \in \{1, \ldots, 6\},
\label{E:RWN_PerV}
\end{equation}
where $k$ identifies the climate state of observation $y_n$, according to the six Cenozoic climate states identified by \cite{westerhold2020} and used in \cite{bhkl2024}; the timing of the transitions between these states is statistically confirmed by \cite{larsen2023}.

In addition to the RWN model, we also fit the integrated random walk plus noise (IWN) model of order $m$, as described in \cite{bhkl2024}. The IWN model replaces the random walk with a higher-order integrated process: the latent level is the $m$-fold integral of a single white-noise innovation entering at the $m$-th derivative, yielding smoother signal extraction at the cost of stronger smoothing assumptions. We consider $m = 2$.

All parameters are estimated by maximum likelihood, using the prediction error decomposition computed via the Kalman filter with diffuse initialization for the non-stationary initial state $\mu_0$.

\subsection{Trivariate random walk plus noise}\label{S:TriRWN}

The central contribution of this paper is the joint modeling of $\delta^{18}$O, $\delta^{13}$C, and $\log(\text{CO}_2)$ in a single trivariate state space model. This extends the bivariate model for $\delta^{18}$O and $\delta^{13}$C developed in \cite{bhkl2024} by adding $\log(\text{CO}_2)$ as a third latent state.

The trivariate state vector is $\boldsymbol{\mu}_{t} = (\mu^{\delta^{18}O}_{t},\, \mu^{\delta^{13}C}_{t},\, \mu^{\log\text{CO}_2}_{t})'$, with dimension 3. The measurement equations are
\begin{align}
y^{\delta^{18}O}_{n} &= c^{\delta^{18}O}_s + \mu^{\delta^{18}O}_{t_n} + \eps^{\delta^{18}O}_{n}, & \eps^{\delta^{18}O}_{n} &\stackrel{indep.}{\sim} \mathsf{N}(0, \sigma^2_{\eps,s,\delta^{18}O}), & s &\in \{1,\ldots,10\}, \notag \\
y^{\delta^{13}C}_{n} &= c^{\delta^{13}C}_s + \mu^{\delta^{13}C}_{t_n} + \eps^{\delta^{13}C}_{n}, & \eps^{\delta^{13}C}_{n} &\stackrel{indep.}{\sim} \mathsf{N}(0, \sigma^2_{\eps,s,\delta^{13}C}), & s &\in \{1,\ldots,10\}, \label{E:TriMeas} \\
y^{\log\text{CO}_2}_{n} &= c_g + \mu^{\log\text{CO}_2}_{t_n} + \eps^{\log\text{CO}_2}_{n}, & \eps^{\log\text{CO}_2}_{n} &\stackrel{indep.}{\sim} \mathsf{N}(0, \sigma_{\eps,g}^2), & g &\in \{1,\ldots,7\}, \notag
\end{align}
where the measurement error variances and intercepts are differentiated by drill site ($s$) for the isotope proxies, as in the bivariate model of \cite{bhkl2024}, and by proxy group ($g$; seven groups) for CO$_2$. The intercepts ($c^{\delta^{18}O}_s$ and $c^{\delta^{13}C}_s$ for the isotopes, $c_g$ for CO$_2$) are the bias correction that places all proxies on a common latent scale; they are normalized to zero at a reference site and at the \texttt{b\_ca} reference group. The measurement errors are mutually independent, both across proxies and across observations.

The transition equations combine a stochastic Brownian innovation with a deterministic, climate-state-dependent Milankovitch forcing of the level:
\begin{align}
\mu^{\delta^{18}O}_{t_n + \Delta t_n} &= \mu^{\delta^{18}O}_{t_n} + b_{11,k}\,\Delta e_{t_n} + b_{21,k}\,\Delta\varepsilon_{t_n} + b_{31,k}\,\Delta\bar\omega_{t_n} + \eta^{\delta^{18}O}_{t_n}, \notag \\
\mu^{\delta^{13}C}_{t_n + \Delta t_n} &= \mu^{\delta^{13}C}_{t_n} + b_{12,k}\,\Delta e_{t_n} + b_{22,k}\,\Delta\varepsilon_{t_n} + b_{32,k}\,\Delta\bar\omega_{t_n} + \eta^{\delta^{13}C}_{t_n}, \label{E:TriTrans} \\
\mu^{\log\text{CO}_2}_{t_n + \Delta t_n} &= \mu^{\log\text{CO}_2}_{t_n} + b_{13,k}\,\Delta e_{t_n} + b_{23,k}\,\Delta\varepsilon_{t_n} + b_{33,k}\,\Delta\bar\omega_{t_n} + \eta^{\log\text{CO}_2}_{t_n}, \notag
\end{align}
where $\Delta e_{t_n}$, $\Delta\varepsilon_{t_n}$, and $\Delta\bar\omega_{t_n}$ are the increments over $[t_n,\, t_n+\Delta t_n]$ of orbital eccentricity $e$, obliquity $\varepsilon$, and the climatic precession index $\bar\omega = e\sin\varpi$ (with $\varpi$ the longitude of perihelion), computed from the La2004 astronomical solution of \cite{laskar2004}, and the coefficient $b_{ij,k}$ measures the sensitivity of proxy $j$ to orbital driver $i \in \{$eccentricity, obliquity, precession$\}$ within climate state $k = k(n)$, that of the interval leaving observation $n$. Allowing the orbital coefficients to depend on the climate state lets the astronomical sensitivity strengthen as continental ice sheets grow (Section~\ref{S:MilankCoeffs}); setting all $b_{ij,k} = 0$ recovers the purely stochastic random walk and constraining them equal across states gives a constant-coefficient forcing, both nested restrictions tested in Section~\ref{S:TriBIC}. The state disturbances $\boldsymbol{\eta}_{t_n} = (\eta^{\delta^{18}O}_{t_n},\, \eta^{\delta^{13}C}_{t_n},\, \eta^{\log\text{CO}_2}_{t_n})'$ follow a trivariate normal distribution:
\begin{equation}
\boldsymbol{\eta}_{t_n} \stackrel{i.i.d.}{\sim} \mathsf{N}(\mathbf{0},\, Q_k \, \Delta t_n), \label{E:TriCov}
\end{equation}
with the $3 \times 3$ covariance rate matrix
\begin{equation}
Q_k = \begin{bmatrix}
\sigma^2_{\eta,\delta^{18}O,k} & \rho_{12,k}\, \sigma_{\eta,\delta^{18}O,k}\, \sigma_{\eta,\delta^{13}C,k} & \rho_{13,k}\, \sigma_{\eta,\delta^{18}O,k}\, \sigma_{\eta,\log\text{CO}_2,k} \\
\rho_{12,k}\, \sigma_{\eta,\delta^{18}O,k}\, \sigma_{\eta,\delta^{13}C,k} & \sigma^2_{\eta,\delta^{13}C,k} & \rho_{23,k}\, \sigma_{\eta,\delta^{13}C,k}\, \sigma_{\eta,\log\text{CO}_2,k} \\
\rho_{13,k}\, \sigma_{\eta,\delta^{18}O,k}\, \sigma_{\eta,\log\text{CO}_2,k} & \rho_{23,k}\, \sigma_{\eta,\delta^{13}C,k}\, \sigma_{\eta,\log\text{CO}_2,k} & \sigma^2_{\eta,\log\text{CO}_2,k}
\end{bmatrix},
\label{E:Qmatrix}
\end{equation}
where $k \in \{1,\ldots,6\}$ indexes the climate state. The matrix $Q_k \Delta t_n$ is the variance of a trivariate Brownian motion increment over the time interval $\Delta t_n$, extending the bivariate Brownian motion covariance derived in \cite{bhkl2024} from two to three dimensions. The three correlation parameters are $\rho_{12,k}$ (between $\delta^{18}$O and $\delta^{13}$C), $\rho_{13,k}$ (between $\delta^{18}$O and $\log\text{CO}_2$), and $\rho_{23,k}$ (between $\delta^{13}$C and $\log\text{CO}_2$).

\subsubsection*{Long-form data layout}

Because the three proxies ($\delta^{18}$O, $\delta^{13}$C, $\log\text{CO}_2$) originate from different data sources with different time stamps, we organize the merged data in \emph{long form}: each row corresponds to one scalar observation of one proxy, and all observations are sorted chronologically. The merged data set contains $N = 54{,}498$ observations: 24{,}259 of $\delta^{18}$O, 23{,}939 of $\delta^{13}$C, and 6{,}300 of $\log(\text{CO}_2)$.

At each observation, the Kalman filter performs a scalar measurement update using only the row of the measurement matrix corresponding to the observed proxy: $Z_n = e_{p}'$, where $e_p$ is the $p$-th standard basis vector of $\R^3$ and $p \in \{1,2,3\}$ identifies the proxy type. When multiple observations share the same time stamp (which occurs frequently for the isotope data and occasionally for CO$_2$), the time increment is $\Delta t_n = 0$, and no state noise is added between successive observations at that time stamp. This treatment of simultaneous observations is identical to the approach in \cite{bhkl2024}.

\subsubsection*{Parameter counts}

In the simplest trivariate specification, all measurement variances are pooled (one per proxy), the transition parameters are constant across climate states, and there is no orbital forcing, giving 9~parameters: three $\sigma_\eps$, three $\sigma_\eta$, and three correlations $\rho_{12}$, $\rho_{13}$, $\rho_{23}$.

In the full specification, measurement variances are differentiated by drill site for the isotope proxies (10~each, corresponding to the 10 DSDP/ODP/IODP sites in the CENOGRID compilation) and by proxy group for CO$_2$ (7), yielding $10 + 10 + 7 = 27$ measurement variance parameters; the bias intercepts add $9 + 9 + 6 = 24$ more. The transition variances and correlations are differentiated by climate state, yielding $3 \times 6 = 18$ state variance and $3 \times 6 = 18$ correlation parameters. The orbital block contributes $3 \times 3 \times 6 = 54$ climate-state-specific coefficients $b_{ij,k}$. The preferred model therefore has $27 + 24 + 18 + 18 + 54 = 141$ parameters; restricting the orbital coefficients to be equal across climate states gives 96, dropping the orbital forcing altogether gives the bias-corrected 87-parameter random walk, and dropping the intercepts as well gives the 63-parameter ``full'' baseline.

\subsubsection*{Parameter transformations}

To ensure that the variance parameters are positive and the correlation parameters lie in $(-1,1)$, we estimate the models in terms of transformed parameters. We parameterize $\log \sigma_\eps$ and $\log \sigma_\eta$ (so that $\sigma_{\eps}^2 = \exp(\theta)^2$ is always positive), and $\text{atanh}(\rho) = \frac{1}{2}\log\frac{1+\rho}{1-\rho}$ for the correlations (so that $\rho = \tanh(\theta)$ is always in $(-1,1)$). All optimization is performed over the unconstrained transformed parameters using the L-BFGS-B algorithm.

\subsection{Trivariate IWN(2) with fixed state variances}\label{S:TriIWN2}

As a signal-smoothing counterpart to the trivariate random walk, we also consider a trivariate pure $m$-fold integrated random walk plus noise of order $m=2$, extending the univariate IWN(2) of Section~\ref{S:UniCO2} to all three proxies jointly. Each latent proxy $p$ now carries a two-dimensional state $(\mu^{p}_{t},\, \nu^{p}_{t})'$ consisting of a level $\mu^p_t$ and a slope $\nu^p_t$. A single white-noise innovation enters at the slope (the highest order), and the level is its integral with no innovation of its own; this is the pure $m$-fold specification that delivers the order-$m$ signal smoothing of \cite{bhkl2024}. The measurement equations~\eqref{E:TriMeas} are unchanged: only the level $\mu^p_{t_n}$ is observed, $Z_n = e_p' \otimes (1,0)$. Cross-proxy dependence is imposed on the slope innovations, whose $3\times 3$ covariance rate matrix takes the same form as $Q_k$ in~\eqref{E:Qmatrix} but now governs the slopes, with per-period correlations $\rho_{12,k}$, $\rho_{13,k}$, $\rho_{23,k}$ and per-period slope-innovation standard deviations $\sigma_{\eta,p,k}$.

For the CO$_2$ measurement equation we use the same seven proxy groups as the random-walk model of Section~\ref{S:TriRWN}, pooling the six stomatal sub-proxies together with liverwort (each represented by very few observations) into a single \emph{stomata} group. This gives seven CO$_2$ measurement groups (\texttt{b\_ca}, boron isotopes, land-plant $\delta^{13}$C, nahcolite, paleosol, phytoplankton, and pooled stomata), each with its own intercept relative to the \texttt{b\_ca} reference.

The fully free trivariate IWN(2), in which all $3\times 6$ slope-innovation variances are estimated jointly with the correlations, is numerically ill-behaved: the sparse CO$_2$ record allows the flexible slope state to extrapolate linearly across the data gaps, so the filter tracks individual observations instead of smoothing, the smoothed CO$_2$ path can diverge, and the likelihood is poorly identified. We therefore \emph{fix} the six per-period slope-innovation variances of each proxy at the estimates from the corresponding univariate IWN(2) specification: for $\delta^{18}$O and $\delta^{13}$C, the drill-site model of \cite{bhkl2024} with site intercepts and per-period state variances; for $\log(\text{CO}_2)$, a univariate IWN(2) with the same proxy-type measurement variances, proxy intercepts, and per-period state variances. Holding these variances fixed pins the smoothing bandwidth at the well-identified univariate values, and we re-estimate jointly only the bias intercepts, the measurement variances, and the per-period slope-innovation correlations (and, in Section~\ref{S:TriIWN2Results}, the orbital coefficients). The model therefore has $24$ intercepts $+\, 27$ measurement variances $+\, 18$ correlations $= 69$ freely estimated parameters.

The full system matrices of both the random-walk and the IWN(2) specifications are spelled out in Appendix~\ref{A:Matrices}.


\section{Univariate results for $\log(\text{CO}_2)$}\label{S:UniResults}

We fit five univariate models of increasing complexity to the $N_{\text{CO}_2} = 6{,}300$ observations of $\log(\text{CO}_2)$. Table~\ref{T:CO2_BIC} reports the number of parameters, maximized log-likelihood, and Bayes information criterion (BIC) for each model.

\begin{table}[htbp]
\caption{\footnotesize Model comparison for univariate $\log(\text{CO}_2)$ models. $P$: number of parameters. $-\ell$: negative log-likelihood. BIC: Bayes information criterion. Smaller BIC is preferred. ``group variances'' indicates per-group measurement variances, each with a group-specific intercept $c_g$ (reference: \texttt{b\_ca}); the seven proxy groups pool the six stomatal sub-proxies and liverwort into a single stomata group. ``states'' indicates per-period state variances. IWN($m$) denotes the integrated random walk of order $m$.\label{T:CO2_BIC}}
\centering
\begin{tabular}{lrrr}\hline
Model & $P$ & $-\ell$ & BIC \\\hline
RWN pooled & 2 & 1493.30 & 3004.09 \\
RWN group variances + intercepts & 14 & 906.69 & 1935.85 \\
RWN states & 7 & 1321.24 & 2703.72 \\
RWN group variances + states + intercepts & 19 & 805.69 & 1777.59 \\
IWN(2) group variances + states + intercepts & 19 & 1148.04 & 2462.30 \\
\hline
\end{tabular}
\end{table}

The BIC-preferred model is the RWN with per-group measurement variances (each with its own intercept) and per-period state variances (19~parameters, BIC = 1777.59). Adding group-specific measurement variances and intercepts improves the BIC by 1068~points relative to the pooled model, and adding period-specific state variances improves it by a further 158~points. The pure $m$-fold IWN(2), with the same number of parameters as the RWN (19), attains a substantially higher BIC ($-\ell = 1148.04$, BIC = 2462.30). The smoothing necessarily worsens the fit and therefore implies higher residual sums of squares, deteriorating the likelihood and the BIC. This smoothing is a deliberate design choice rather than a deficiency of the model, and we include the IWN(2) in the table only for completeness. We consider $m=2$ only.

\begin{figure}[htbp]
\centering
\includegraphics[width=\textwidth]{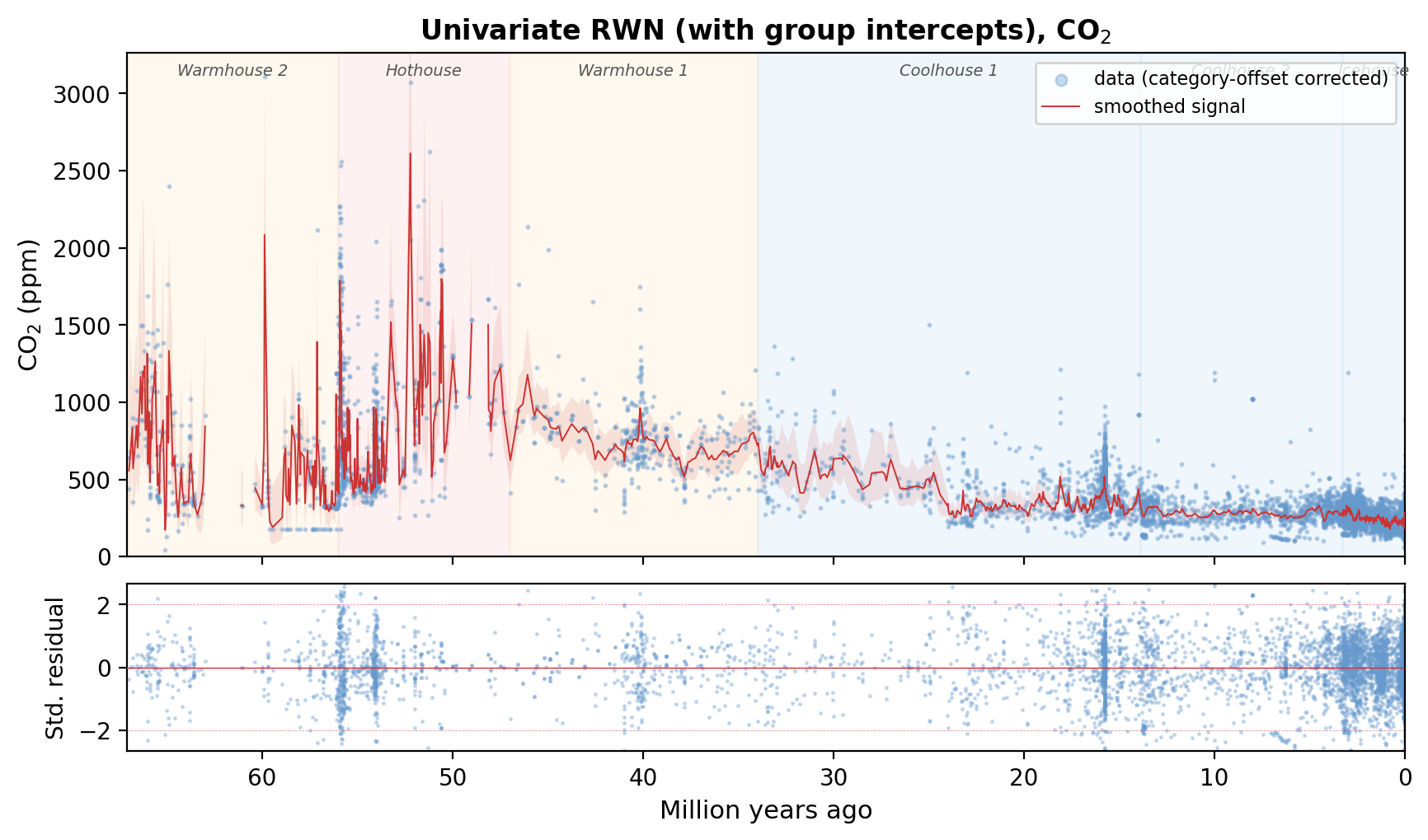}
\caption{\footnotesize Univariate RWN model for CO$_2$ (per-group measurement variances and intercepts, per-period state variances). Top panel: CO$_2$ observations with their estimated group offsets $\hat c_g$ removed (blue dots) and smoothed state (red line) with 95\% confidence band, displayed in ppm. The smoothed level and band are omitted across CO$_2$ data gaps longer than $0.5$~Myr, where the signal is not constrained by observations. Bottom panel: standardized prediction residuals. The background shading indicates the six climate states of \cite{westerhold2020}.\label{F:CO2_uni}}
\end{figure}

Figure~\ref{F:CO2_uni} shows the data and the smoothed state from the preferred univariate RWN model, with CO$_2$ concentrations displayed in ppm. The smoothed signal recovers the expected Cenozoic CO$_2$ history: high concentrations during the early Eocene ($\sim$1000--2000~ppm), a pronounced drawdown around the Eocene--Oligocene transition ($\sim$34~Ma), broadly stable concentrations through the Oligocene and Miocene ($\sim$300--600~ppm), and low concentrations during the Plio-Pleistocene ($\sim$200--300~ppm). The 95\% confidence band is widest in the earliest part of the record, where data are sparsest. The standardized prediction residuals (lower panel) show no obvious trends, though their variance is heterogeneous across time.

\begin{figure}[htbp]
\centering
\includegraphics[width=\textwidth]{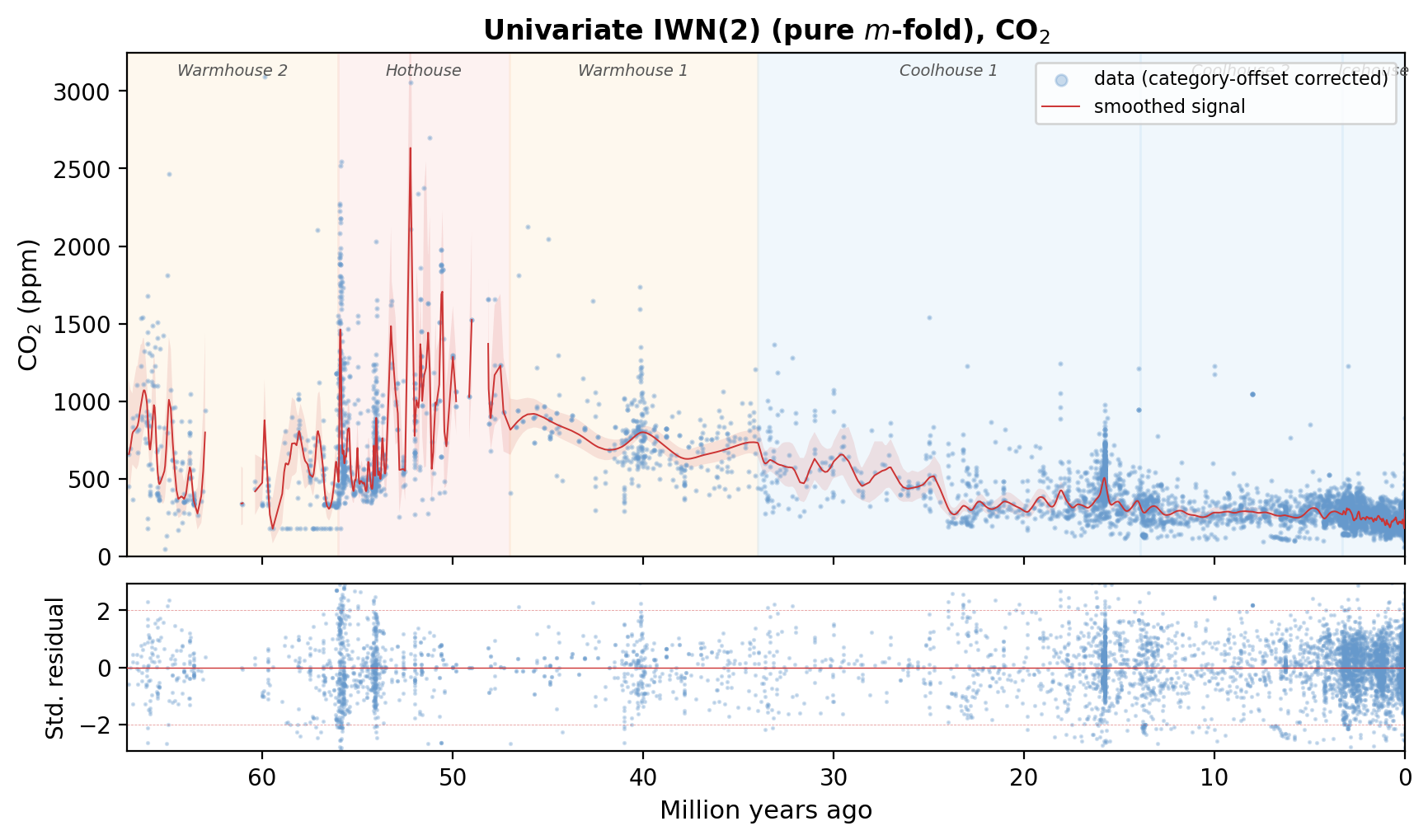}
\caption{\footnotesize Univariate IWN(2) model for CO$_2$ (pure $m$-fold, per-group measurement variances and intercepts, per-period state variances). Top panel: CO$_2$ observations with their estimated group offsets $\hat c_g$ removed (blue dots) and smoothed state (red line) with 95\% confidence band, displayed in ppm. The smoothed level and band are omitted across CO$_2$ data gaps longer than $0.5$~Myr, where the integrated random walk is not constrained by observations. Bottom panel: standardized prediction residuals. Compared to the RWN in Figure~\ref{F:CO2_uni}, the IWN(2) produces a smoother signal that attenuates short-lived excursions.\label{F:CO2_iwn2}}
\end{figure}

For comparison, Figure~\ref{F:CO2_iwn2} shows the smoothed signal from the IWN(2) model with the same measurement and state variance specification. The IWN(2) signal is visibly smoother than the RWN signal in Figure~\ref{F:CO2_uni}, attenuating several of the sharper features, particularly the transient CO$_2$ spikes in the Eocene and the short-lived excursions in the Icehouse period. This additional smoothing is a deliberate design choice rather than a deficiency of the model; we present the IWN(2) signal here only for completeness.

\subsection{Measurement noise by proxy group}\label{S:CO2_eps}

Table~\ref{T:CO2_eps} reports the maximum likelihood estimates of the per-group intercepts $c_g$ and measurement standard deviations $\hat\sigma_{\eps,g}$ for the RWN and IWN(2) models. The intercepts absorb systematic level offsets between proxy groups, measured relative to the \texttt{b\_ca} reference: the boron-isotope and nahcolite groups sit about $0.14$ and $0.93$ in $\log(\text{CO}_2)$ above the reference, while the pooled stomata group sits about $0.04$ below it. Conditional on these offsets, the residual measurement noise still varies substantially across groups and is robust across the two specifications. The nahcolite and paleosol proxies carry the largest residual noise ($\hat\sigma_{\eps} \approx 0.5$--$0.7$), consistent with their providing only loose constraints on CO$_2$; the nahcolite proxy, in particular, brackets CO$_2$ between bounds rather than yielding a point estimate \citep{jagniecki2015}. The boron isotopes ($\hat\sigma_{\eps} \approx 0.21$--$0.23$) and land plant $\delta^{13}$C ($\hat\sigma_{\eps} \approx 0.22$--$0.23$) are among the most precise of the data-rich groups, while phytoplankton ($\hat\sigma_{\eps} \approx 0.35$--$0.36$) is somewhat noisier. The IWN(2) generally estimates slightly larger measurement noise, absorbing into the measurement equation the high-frequency variation that the smoother latent state cannot track.

\begin{table}[htbp]
\caption{\footnotesize Maximum likelihood estimates of per-group intercepts $c_g$ and measurement standard deviations $\hat\sigma_{\eps,g}$ for the RWN and (pure $m$-fold) IWN(2) models with per-group measurement variances and per-period state variances applied to $\log(\text{CO}_2)$. The seven proxy groups pool the six stomatal sub-proxies and liverwort into a single stomata group; \texttt{b\_ca} is the intercept reference ($c_g=0$). Standard errors in parentheses: directly for the intercepts, and by the delta method (from the Hessian-based standard error of the estimated variance) for the measurement standard deviations.\label{T:CO2_eps}}
\centering
\footnotesize
\begin{tabular}{lrrr}\hline
Group & No.\ obs. & RWN & IWN(2) \\\hline
$c_g$ & & & \\
b\_ca & 277 & 0 (ref) & 0 (ref) \\
boron\_isotopes & 2{,}111 & 0.136 $(0.013)$ & 0.141 $(0.013)$ \\
land\_plant\_d13c & 1{,}799 & -0.037 $(0.014)$ & -0.047 $(0.014)$ \\
nahcolite & 4 & 0.928 $(0.408)$ & 0.847 $(0.411)$ \\
paleosol & 319 & -0.020 $(0.034)$ & -0.049 $(0.037)$ \\
phytoplankton & 1{,}201 & 0.030 $(0.016)$ & 0.030 $(0.016)$ \\
stomata (pooled) & 589 & -0.043 $(0.023)$ & -0.070 $(0.023)$ \\
$\hat\sigma_{\eps,g}$ & & & \\
b\_ca & 277 & 0.140 $(0.009)$ & 0.159 $(0.009)$ \\
boron\_isotopes & 2{,}111 & 0.211 $(0.005)$ & 0.233 $(0.004)$ \\
land\_plant\_d13c & 1{,}799 & 0.216 $(0.004)$ & 0.229 $(0.004)$ \\
nahcolite & 4 & 0.607 $(0.335)$ & 0.686 $(0.329)$ \\
paleosol & 319 & 0.513 $(0.026)$ & 0.588 $(0.025)$ \\
phytoplankton & 1{,}201 & 0.347 $(0.008)$ & 0.356 $(0.008)$ \\
stomata (pooled) & 589 & 0.285 $(0.012)$ & 0.335 $(0.012)$ \\
\hline
\end{tabular}
\end{table}

\subsection{State variance by climate state}\label{S:CO2_eta}

Table~\ref{T:CO2_eta} reports the maximum likelihood estimates of the per-period state standard deviations $\hat\sigma_{\eta,k}$ for the RWN and IWN(2) models. For the RWN, the estimates reveal that the latent CO$_2$ signal was more volatile during the early Cenozoic warm periods than during the later cool periods. The Warmhouse~2 and Hothouse periods have level-innovation standard deviations of 2.58 and 2.00, respectively, whereas the later periods range between 0.28 and 0.72, the Icehouse rate of 0.72 reflecting the well-documented glacial--interglacial CO$_2$ cycles. The IWN(2), whose single state driver is the slope (highest-order) innovation, yields markedly larger and more erratic estimates (e.g.\ 10.0 and 38.0 in the two warmest periods). These are not directly comparable to the RWN level-innovation rates: the slope-innovation variance enters the level only after a double time integration, so generating comparable level variation from finely spaced data requires a much larger rate.

\begin{table}[htbp]
\caption{\footnotesize Maximum likelihood estimates of per-period state standard deviations $\hat\sigma_{\eta,k}$ for the RWN and (pure $m$-fold) IWN(2) models with per-group measurement variances (and group intercepts) and per-period state variances applied to $\log(\text{CO}_2)$. Climate states as defined in \cite{westerhold2020}. For the RWN, $\sigma_{\eta,k}$ is the level-innovation standard deviation; for IWN(2) it is the standard deviation of the highest-order (slope) innovation, the sole state driver. The IWN(2) values are larger because the slope-innovation rate carries a $\Delta t^{-(2m-1)}$ scaling (see text). Standard errors in parentheses (delta method, from the Hessian-based standard error of the estimated variance).\label{T:CO2_eta}}
\centering
\footnotesize
\begin{tabular}{llrr}\hline
$k$ & Climate state & RWN & IWN(2) \\\hline
1 & Warmhouse 2 (67--56 Ma) & 2.575 $(0.340)$ & 10.013 $(2.809)$ \\
2 & Hothouse (56--47 Ma) & 2.004 $(0.213)$ & 38.043 $(5.343)$ \\
3 & Warmhouse 1 (47--34 Ma) & 0.328 $(0.098)$ & 0.172 $(0.083)$ \\
4 & Coolhouse 1 (34--13.9 Ma) & 0.597 $(0.113)$ & 1.953 $(0.399)$ \\
5 & Coolhouse 2 (13.9--3.3 Ma) & 0.277 $(0.053)$ & 1.107 $(0.314)$ \\
6 & Icehouse (3.3--0 Ma) & 0.724 $(0.076)$ & 23.468 $(3.436)$ \\
\hline
\end{tabular}
\end{table}


\section{Trivariate results}\label{S:TriResults}

\subsection{Model comparison}\label{S:TriBIC}

We fit trivariate models of increasing complexity to the merged long-form data set of $N = 54{,}498$ observations. Table~\ref{T:Tri_BIC} reports the Bayes information criterion along a ladder of nested specifications, from a simple pooled random walk plus noise to the preferred model with climate-state-dependent innovation covariance and period-dependent Milankovitch forcing.

\begin{table}[htbp]
\caption{\footnotesize Model comparison for trivariate $(\delta^{18}\text{O},\, \delta^{13}\text{C},\, \log\text{CO}_2)$ models of increasing complexity. ``simple'': one measurement and one state variance per proxy, pooled across climate states. ``full'': measurement variances differentiated by drill site ($\delta^{18}$O, $\delta^{13}$C) and proxy group (CO$_2$), state variances and correlations differentiated by climate state. ``+ bias correction'': adding site and group intercepts. The last two rows add the Milankovitch forcing of the level increments, first with climate-state-invariant coefficients and then with period-dependent coefficients (the preferred model). $P$: number of parameters. $-\ell$: negative log-likelihood. BIC: Bayes information criterion (smaller preferred). $\Delta$BIC and LR: change relative to the row above; df: added parameters.\label{T:Tri_BIC}}
\centering\footnotesize
\begin{tabular}{lrrrrrr}\hline
Model & $P$ & $-\ell$ & BIC & $\Delta$BIC & LR & df \\\hline
Trivariate RWN, simple & 9 & $-8{,}744.65$ & $-17{,}391.15$ & & & \\
Trivariate RWN, full & 63 & $-19{,}940.08$ & $-39{,}193.10$ & & & \\
Trivariate RWN, full + bias correction & 87 & $-20{,}722.17$ & $-40{,}495.53$ & & & \\
\quad + Milankovitch forcing & 96 & $-20{,}841.01$ & $-40{,}635.06$ & $-139.53$ & $237.68$ & 9 \\
\quad + period-dependent Milank.\ forcing & 141 & $-21{,}175.11$ & $-40{,}812.48$ & $-177.42$ & $668.19$ & 45 \\
\hline
\end{tabular}
\end{table}

Differentiating the measurement and state variances by site, group, and climate state improves the BIC dramatically, from $-17{,}391$ for the simple model to $-39{,}193$ for the full model. Adding the site and group intercepts (the bias correction that places all proxies on a common latent scale, as in the univariate CO$_2$ analysis of Section~\ref{S:UniResults} and the bivariate isotope analysis of \cite{bhkl2024}) improves the BIC by a further $1{,}303$ points to $-40{,}496$ (a likelihood-ratio statistic of $1{,}564$ on 24 degrees of freedom). As in the univariate CO$_2$ analysis, the random-walk dynamics capture the signal-to-noise structure of the merged record efficiently; the higher-order integrated random walk, a signal-smoothing device, is reported in Section~\ref{S:TriIWN2Results}.

The model of Section~\ref{S:TriRWN} carries the orbital forcing through the climate-state-dependent coefficient matrix $B_{k(n)}$ in the transition equation~\eqref{E:TriTrans}, which lets the astronomical sensitivity of each proxy differ across the six climate states ($3\times3\times6 = 54$ coefficients). We assess this orbital block through two nested restrictions. Setting all $b_{ij,k}=0$ recovers the bias-corrected baseline RWN; imposing instead a single, climate-state-invariant matrix $B=(b_{ij})$ gives the constant-coefficient specification,

\begin{align}
\mu^{\delta^{18}O}_{t_n + \Delta t_n} &= \mu^{\delta^{18}O}_{t_n} + b_{11}\,\Delta e_{t_n} + b_{21}\,\Delta\varepsilon_{t_n} + b_{31}\,\Delta\bar\omega_{t_n} + \eta^{\delta^{18}O}_{t_n}, \notag \\
\mu^{\delta^{13}C}_{t_n + \Delta t_n} &= \mu^{\delta^{13}C}_{t_n} + b_{12}\,\Delta e_{t_n} + b_{22}\,\Delta\varepsilon_{t_n} + b_{32}\,\Delta\bar\omega_{t_n} + \eta^{\delta^{13}C}_{t_n}, \label{E:MilankTrans} \\
\mu^{\log\text{CO}_2}_{t_n + \Delta t_n} &= \mu^{\log\text{CO}_2}_{t_n} + b_{13}\,\Delta e_{t_n} + b_{23}\,\Delta\varepsilon_{t_n} + b_{33}\,\Delta\bar\omega_{t_n} + \eta^{\log\text{CO}_2}_{t_n}, \notag
\end{align}

in which the rows of $B$ index the three orbital components and the columns the three proxies (9 coefficients), and the stochastic innovations $\boldsymbol{\eta}_{t_n}$ retain the covariance structure~\eqref{E:TriCov}. A single, time-invariant $B$ is physically restrictive, because the feedbacks that translate insolation changes into the recorded signals (most importantly the ice-albedo feedback) operate strongly only once substantial continental ice sheets exist.

Adding constant orbital coefficients improves the BIC of the bias-corrected RWN by 140~units, with a likelihood-ratio statistic of $\text{LR}=237.7$ on 9~degrees of freedom ($p\approx3\times10^{-46}$, $\chi^2_9$), decisively rejecting the null of no orbital forcing. Letting the coefficients differ across climate states improves the fit substantially further: the BIC falls by an additional 177~units, and the likelihood-ratio statistic for equality across states is $\text{LR}=668.2$ on 45~degrees of freedom, far beyond the $\chi^2_{45}$ critical value of $61.7$. A single coefficient matrix must compromise between climate states whose response to orbital forcing differs in kind, averaging the large sensitivities of the glaciated late Cenozoic together with the near-absence of any response in the ice-free early Cenozoic (Section~\ref{S:MilankCoeffs}). We therefore adopt the bias-corrected trivariate RWN with period-dependent Milankovitch forcing (141~parameters) as our preferred specification, and report it throughout.

\subsection{Orbital forcing coefficients}\label{S:MilankCoeffs}

\begin{figure}[htbp]
\centering
\includegraphics[width=\textwidth]{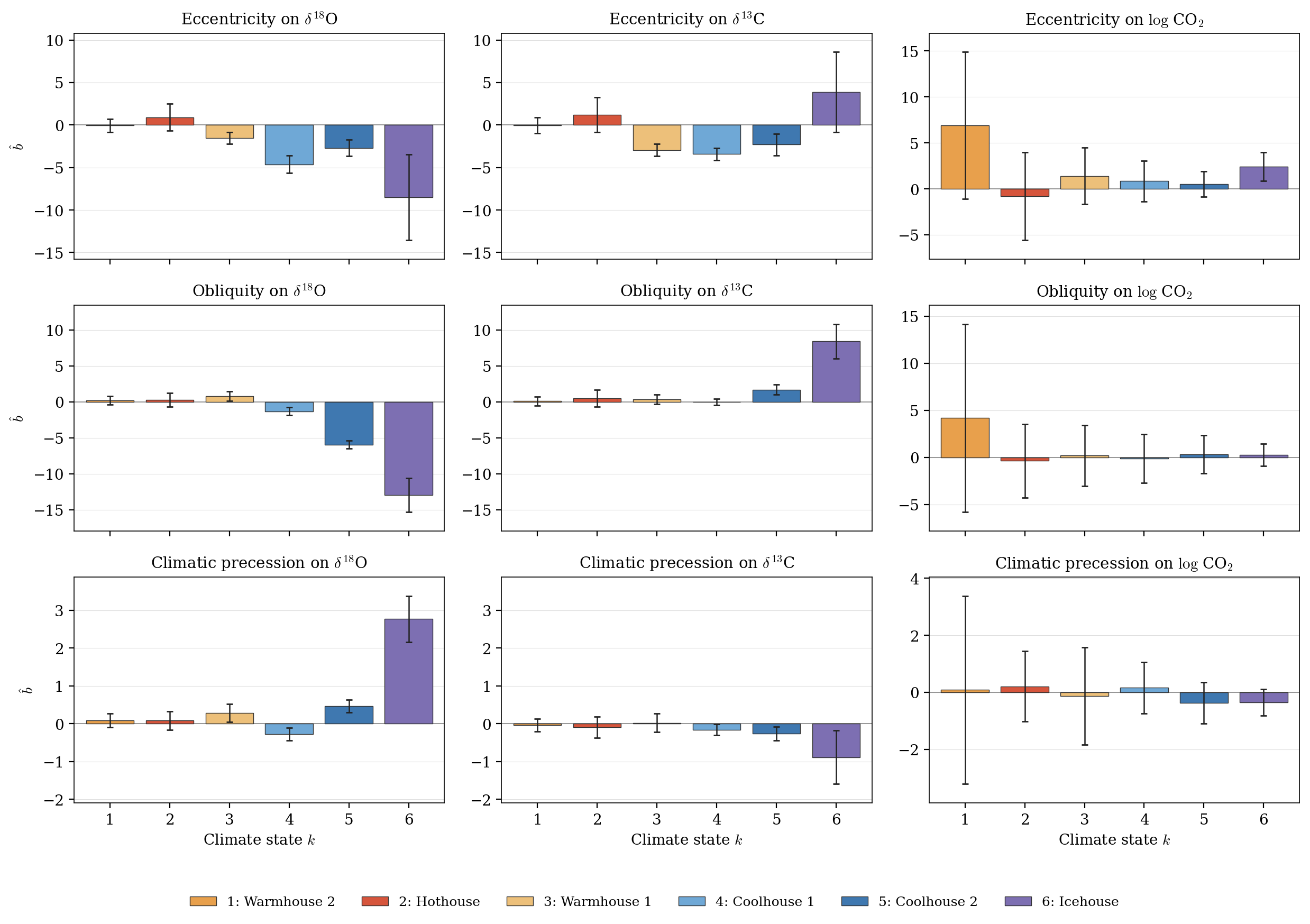}
\caption{\footnotesize Estimated period-dependent Milankovitch coefficients $\hat b_{ij,k}$ from the preferred trivariate RWN model (141~parameters), by climate state. Rows are the three orbital variables (eccentricity, obliquity, climatic precession); columns are the three proxies ($\delta^{18}$O, $\delta^{13}$C, $\log\text{CO}_2$). Each bar is the coefficient within one climate state, with $\pm 2$ standard-error whiskers; the abscissa labels 1 through 6 are the six Westerhold climate states (Table~\ref{T:MilankOrb} gives the calendar intervals and the numerical values). The $\delta^{18}$O and $\delta^{13}$C columns, both in permil, share a common vertical scale within each row; the $\log\text{CO}_2$ column, in different units and with much larger standard errors, is drawn on its own scale.\label{F:MilankOrb}}
\end{figure}

Table~\ref{T:MilankOrb} reports the estimated period-dependent orbital coefficients $\hat{b}_{ij,k}$ from the preferred trivariate RWN model with climate-state-specific Milankovitch forcing, with standard errors computed from the numerical Hessian of the log-likelihood. Figure~\ref{F:MilankOrb} displays the same estimates as climate-state bar charts, which makes the progression across states immediate to read.

\begin{table}[htbp]
\caption{\footnotesize Estimated period-dependent orbital forcing coefficients $\hat{b}_{ij,k}$ from the trivariate RWN + Milankovitch model with climate-state-specific coefficients (141~parameters). Each coefficient multiplies the increment of the orbital variable over an inter-observation interval in the level transition of the respective proxy, separately within each of the six climate states. Standard errors in parentheses, computed by inverting the numerical Hessian. Significance: ${}^{***}p<0.01$, ${}^{**}p<0.05$.\label{T:MilankOrb}}
\centering\footnotesize
\begin{tabular}{llrrr}\hline
Climate state & Orbital variable & $\delta^{18}$O & $\delta^{13}$C & $\log\text{CO}_2$ \\\hline
Warmhouse 2 (67--56 Ma)    & Eccentricity       & $-0.051$ & $-0.039$ & $+7.208$ \\
                           &                    & $(0.391)$ & $(0.454)$ & $(4.057)$ \\
                           & Obliquity          & $+0.193$ & $+0.109$ & $+3.497$ \\
                           &                    & $(0.301)$ & $(0.308)$ & $(5.066)$ \\
                           & Clim.\ precession  & $+0.079$ & $-0.041$ & $+0.255$ \\
                           &                    & $(0.091)$ & $(0.086)$ & $(1.660)$ \\[3pt]
Hothouse (56--47 Ma)       & Eccentricity       & $+0.934$ & $+1.230$ & $-0.779$ \\
                           &                    & $(0.794)$ & $(1.039)$ & $(2.338)$ \\
                           & Obliquity          & $+0.270$ & $+0.505$ & $-0.107$ \\
                           &                    & $(0.479)$ & $(0.595)$ & $(1.902)$ \\
                           & Clim.\ precession  & $+0.080$ & $-0.099$ & $+0.196$ \\
                           &                    & $(0.122)$ & $(0.142)$ & $(0.610)$ \\[3pt]
Warmhouse 1 (47--34 Ma)    & Eccentricity       & $-1.516^{***}$ & $-2.940^{***}$ & $+0.902$ \\
                           &                    & $(0.342)$ & $(0.358)$ & $(1.424)$ \\
                           & Obliquity          & $+0.816^{**}$ & $+0.363$ & $+0.694$ \\
                           &                    & $(0.324)$ & $(0.328)$ & $(1.598)$ \\
                           & Clim.\ precession  & $+0.280^{**}$ & $+0.021$ & $+0.070$ \\
                           &                    & $(0.121)$ & $(0.121)$ & $(0.811)$ \\[3pt]
Coolhouse 1 (34--13.9 Ma)  & Eccentricity       & $-4.614^{***}$ & $-3.417^{***}$ & $+0.720$ \\
                           &                    & $(0.504)$ & $(0.358)$ & $(1.081)$ \\
                           & Obliquity          & $-1.306^{***}$ & $-0.018$ & $+0.051$ \\
                           &                    & $(0.277)$ & $(0.219)$ & $(1.257)$ \\
                           & Clim.\ precession  & $-0.274^{***}$ & $-0.164^{**}$ & $+0.310$ \\
                           &                    & $(0.084)$ & $(0.073)$ & $(0.445)$ \\[3pt]
Coolhouse 2 (13.9--3.3 Ma) & Eccentricity       & $-2.682^{***}$ & $-2.278^{***}$ & $+0.549$ \\
                           &                    & $(0.495)$ & $(0.637)$ & $(0.692)$ \\
                           & Obliquity          & $-5.950^{***}$ & $+1.714^{***}$ & $+0.340$ \\
                           &                    & $(0.281)$ & $(0.348)$ & $(1.008)$ \\
                           & Clim.\ precession  & $+0.462^{***}$ & $-0.263^{***}$ & $-0.352$ \\
                           &                    & $(0.081)$ & $(0.092)$ & $(0.363)$ \\[3pt]
Icehouse (3.3--0 Ma)       & Eccentricity       & $-8.746^{***}$ & $+4.045$ & $+2.448^{***}$ \\
                           &                    & $(2.512)$ & $(2.356)$ & $(0.771)$ \\
                           & Obliquity          & $-13.013^{***}$ & $+8.499^{***}$ & $+0.282$ \\
                           &                    & $(1.186)$ & $(1.205)$ & $(0.585)$ \\
                           & Clim.\ precession  & $+2.767^{***}$ & $-0.889^{**}$ & $-0.347$ \\
                           &                    & $(0.305)$ & $(0.354)$ & $(0.235)$ \\
\hline
\end{tabular}
\end{table}

The coefficients reveal a pronounced dependence on climate state that the constant-coefficient model conceals by averaging over it. In the warm early Cenozoic (the Warmhouse~2 and Hothouse states, 67--47~Ma), essentially none of the orbital coefficients are statistically significant: while the Earth was largely ice-free, the recorded isotope and CO$_2$ signals carry little systematic imprint of orbital forcing. From the Warmhouse~1 onward the orbital signal emerges and then strengthens monotonically into the Icehouse. \emph{Eccentricity} acts strongly and significantly on both isotopes from $\sim$47~Ma on, reaching $\hat{b} = -4.61$ on $\delta^{18}$O in the Coolhouse~1 ($t = -9.2$); higher eccentricity is associated with lower (warmer, less glaciated) $\delta^{18}$O and lower $\delta^{13}$C. The \emph{obliquity} response on $\delta^{18}$O is the most striking: negligible and insignificant in the warm states, it grows to $-1.31$ in the Coolhouse~1, $-5.95$ in the Coolhouse~2 ($t = -21$), and $-13.01$ in the Icehouse ($t = -11$). This intensifying obliquity control is exactly what one expects physically, mirroring the well-documented obliquity pacing of late-Cenozoic glacial cycles through the ice-albedo feedback, which can only operate once large ice sheets are present. \emph{Climatic precession}, though an order of magnitude smaller, likewise becomes significant on the isotopes in the cooler states.

\begin{figure}[htbp]
\centering
\includegraphics[width=\textwidth]{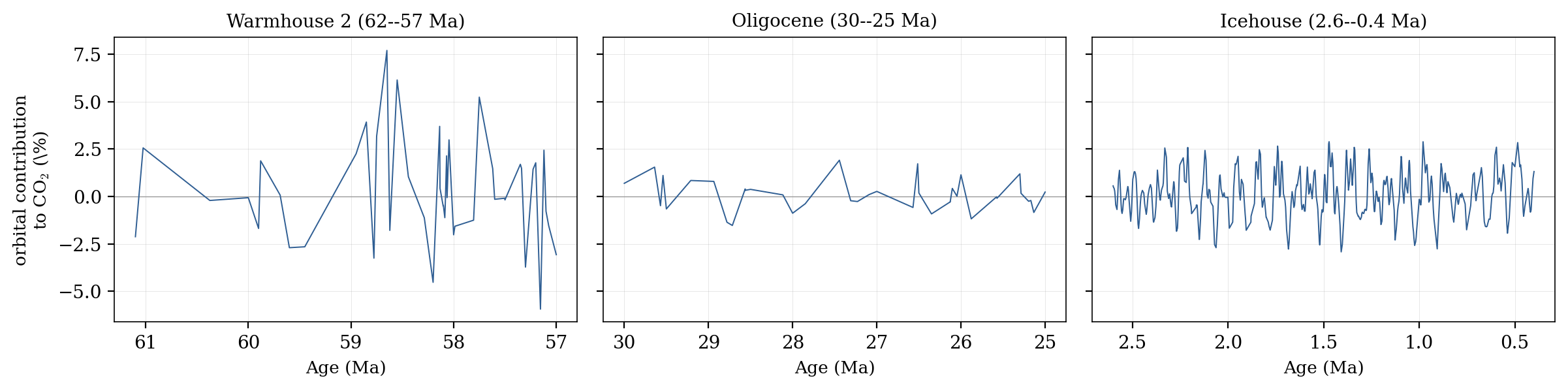}
\caption{\footnotesize Orbital contribution to the smoothed CO$_2$ level: the percentage difference $100\,[\exp(\hat\mu^{\,\text{with}} - \hat\mu^{\,\text{without}})-1]$ between the Kalman-smoothed $\log(\text{CO}_2)$ level of the preferred period-dependent Milankovitch model and the same fitted model with the orbital coefficients set to zero, in three windows (the estimated variances and intercepts are held fixed). The contribution is small throughout, at most a few percent, consistent with the weak and mostly insignificant CO$_2$ orbital coefficients in Table~\ref{T:MilankOrb}, and is most pronounced in the densely sampled Icehouse. The abscissa is age in millions of years before present.\label{F:OrbContrib}}
\end{figure}

By contrast, the $\log\text{CO}_2$ coefficients are weak and mostly insignificant in every climate state (only eccentricity in the Icehouse is significant). This plausibly reflects the limited temporal resolution of the CO$_2$ proxy data: with only $\sim$6{,}300 observations spread over 67~million years, the orbital-band cycles (particularly the $\sim$21~kyr precession and $\sim$41~kyr obliquity bands) are difficult to resolve in CO$_2$, whereas the densely sampled benthic isotopes capture them well, especially on the longer 100 and 405~kyr eccentricity timescales. Figure~\ref{F:OrbContrib} makes this explicit, plotting the orbital contribution to the smoothed CO$_2$ level (the difference between the smoothed path with and without the orbital term, the estimated variances and intercepts held fixed); it is at most a few percent in every window.

\subsection{Correlations by climate state}\label{S:TriCorr}

Table~\ref{T:MilankCorr} reports the three pairwise correlation estimates $\hat\rho_{12,k}$, $\hat\rho_{13,k}$, $\hat\rho_{23,k}$ for each of the six climate states. The correlations between $\delta^{18}$O and $\delta^{13}$C ($\hat\rho_{12,k}$) reproduce the pattern found in the bivariate analysis of \cite{bhkl2024}: strongly positive in the warm periods ($+0.81$ to $+0.97$), weakening through the Coolhouse ($+0.60$ then $+0.21$), and reversing to strongly negative in the Icehouse ($-0.77$). This consistency validates the trivariate filter.

The correlations involving $\log(\text{CO}_2)$ reveal new patterns. The $\delta^{18}$O--$\log\text{CO}_2$ correlation ($\hat\rho_{13,k}$) is small in the warm periods, but turns negative in the Coolhouse~2 ($-0.28$) and Icehouse ($-0.41$) periods. Since declining $\delta^{18}$O values indicate warming (the $\delta^{18}$O axis is conventionally inverted; benthic $\delta^{18}$O reflects deep-ocean temperature together with global ice volume, with the standard piecewise conversion to temperature given by \citealp{hansen2013climate}), a negative correlation between $\delta^{18}$O and $\log\text{CO}_2$ is consistent with the expected greenhouse effect: rising CO$_2$ accompanies declining $\delta^{18}$O (warming and, after $\sim$34~Ma, declining ice volume). The emergence of this negative correlation specifically in the ice-age periods suggests that the coupling of CO$_2$ to temperature and ice volume strengthens when ice sheets are present to amplify the feedback.

The $\delta^{13}$C--$\log\text{CO}_2$ correlation ($\hat\rho_{23,k}$) is weakly negative in the Hothouse ($-0.15$) and Coolhouse~2 ($-0.57$) periods and weakly positive in the Icehouse ($+0.23$). The negative correlation in Coolhouse~2 is the strongest of all $\delta^{13}$C--$\log\text{CO}_2$ correlations, potentially reflecting changes in ocean carbon cycling during the late Miocene.

\begin{table}[htbp]
\caption{\footnotesize Maximum likelihood estimates of pairwise correlations $\hat\rho_k$ from the period-dependent RWN + Milankovitch model (141~parameters), by climate state. 95\% confidence intervals (in brackets) are obtained from the Hessian-based standard errors of the estimated $\operatorname{atanh}\hat\rho_k$ via the Fisher-$z$ transform. Climate states as defined in \cite{westerhold2020}.\label{T:MilankCorr}}
\centering\footnotesize
\begin{tabular}{lrrr}\hline
Climate state & $\hat\rho_{12,k}$ & $\hat\rho_{13,k}$ & $\hat\rho_{23,k}$ \\
 & ($\delta^{18}$O, $\delta^{13}$C) & ($\delta^{18}$O, $\log\text{CO}_2$) & ($\delta^{13}$C, $\log\text{CO}_2$) \\\hline
Warmhouse 2 (67--56 Ma)    & $+0.807$ & $+0.149$ & $+0.106$ \\
                           & $[+0.79,\,+0.82]$ & $[-0.21,\,+0.47]$ & $[-0.22,\,+0.41]$ \\[3pt]
Hothouse (56--47 Ma)       & $+0.966$ & $-0.143$ & $-0.146$ \\
                           & $[+0.97,\,+0.97]$ & $[-0.32,\,+0.04]$ & $[-0.31,\,+0.03]$ \\[3pt]
Warmhouse 1 (47--34 Ma)    & $+0.811$ & $+0.231$ & $+0.305$ \\
                           & $[+0.79,\,+0.83]$ & $[-0.29,\,+0.65]$ & $[-0.24,\,+0.70]$ \\[3pt]
Coolhouse 1 (34--13.9 Ma)  & $+0.599$ & $+0.253$ & $-0.094$ \\
                           & $[+0.56,\,+0.63]$ & $[-0.18,\,+0.61]$ & $[-0.50,\,+0.34]$ \\[3pt]
Coolhouse 2 (13.9--3.3 Ma) & $+0.211$ & $-0.275$ & $-0.570$ \\
                           & $[+0.13,\,+0.29]$ & $[-0.64,\,+0.19]$ & $[-0.80,\,-0.20]$ \\[3pt]
Icehouse (3.3--0 Ma)       & $-0.769$ & $-0.407$ & $+0.229$ \\
                           & $[-0.80,\,-0.74]$ & $[-0.56,\,-0.22]$ & $[-0.06,\,+0.48]$ \\
\hline
\end{tabular}
\end{table}

\subsection{State variance rates by climate state}\label{S:TriEta}

Table~\ref{T:MilankEta} reports the maximum likelihood estimates of the per-period state standard deviations $\hat\sigma_{\eta,k}$ from the preferred period-dependent RWN + Milankovitch model, for each proxy and climate state.

\begin{table}[htbp]
\caption{\footnotesize Maximum likelihood estimates of per-period state standard deviations $\hat\sigma_{\eta,k}$ from the period-dependent RWN + Milankovitch model (141~parameters), by climate state and proxy. Standard errors in parentheses, obtained from the Hessian-based standard errors of the estimated log-variances by the delta method.\label{T:MilankEta}}
\centering\footnotesize
\begin{tabular}{lrrr}\hline
Climate state & $\hat\sigma_{\eta,k}^{\delta^{18}O}$ & $\hat\sigma_{\eta,k}^{\delta^{13}C}$ & $\hat\sigma_{\eta,k}^{\log\text{CO}_2}$ \\\hline
Warmhouse 2 (67--56 Ma)    & 0.692 & 0.855 & 2.473 \\
                           & $(0.015)$ & $(0.022)$ & $(2.104)$ \\[3pt]
Hothouse (56--47 Ma)       & 1.504 & 2.000 & 1.944 \\
                           & $(0.103)$ & $(0.215)$ & $(0.792)$ \\[3pt]
Warmhouse 1 (47--34 Ma)    & 0.548 & 0.595 & 0.346 \\
                           & $(0.011)$ & $(0.013)$ & $(0.014)$ \\[3pt]
Coolhouse 1 (34--13.9 Ma)  & 1.341 & 0.901 & 0.638 \\
                           & $(0.070)$ & $(0.027)$ & $(0.053)$ \\[3pt]
Coolhouse 2 (13.9--3.3 Ma) & 0.976 & 1.273 & 0.309 \\
                           & $(0.037)$ & $(0.065)$ & $(0.006)$ \\[3pt]
Icehouse (3.3--0 Ma)       & 2.874 & 2.622 & 0.625 \\
                           & $(0.825)$ & $(0.886)$ & $(0.026)$ \\
\hline
\end{tabular}
\end{table}

The $\delta^{18}$O and $\delta^{13}$C state variance rates are broadly consistent with the bivariate results of \cite{bhkl2024}, with the highest values in the Icehouse (driven by large glacial--interglacial swings) and the Hothouse (marked by hyperthermal events). The $\log(\text{CO}_2)$ state variance pattern differs markedly: it is highest in the Warmhouse~2 and Hothouse periods ($\hat\sigma_\eta = 2.47$ and $1.94$, respectively) and much lower from the Warmhouse~1 onward ($\hat\sigma_\eta$ between $0.31$ and $0.64$). This indicates that the latent CO$_2$ signal underwent large and rapid changes during the early Cenozoic greenhouse world, but has been comparatively stable (relative to the mean level) since the Eocene--Oligocene transition at 34~Ma.

The contrast between the isotope and CO$_2$ volatility patterns is noteworthy: whereas $\delta^{18}$O variability increased toward the present (peaking in the Icehouse), CO$_2$ variability decreased. This asymmetry may reflect the growing role of ice-sheet dynamics as a source of $\delta^{18}$O variability in the later Cenozoic, a mechanism that does not directly amplify CO$_2$ fluctuations on the same timescales.

In the cool states where the orbital forcing is strongest, the state standard deviations decline relative to the orbital-off restriction (for example, the $\delta^{18}$O innovation standard deviation in the Icehouse falls from $3.39$ to $2.87$, and in the Coolhouse~2 from $1.32$ to $0.98$): the now explicitly resolved, climate-state-specific orbital forcing accounts for part of the variability that the purely stochastic baseline absorbs into the state.

\subsection{Smoothed states}\label{S:MilankFigures}

\begin{figure}[htbp]
\centering
\includegraphics[width=\textwidth]{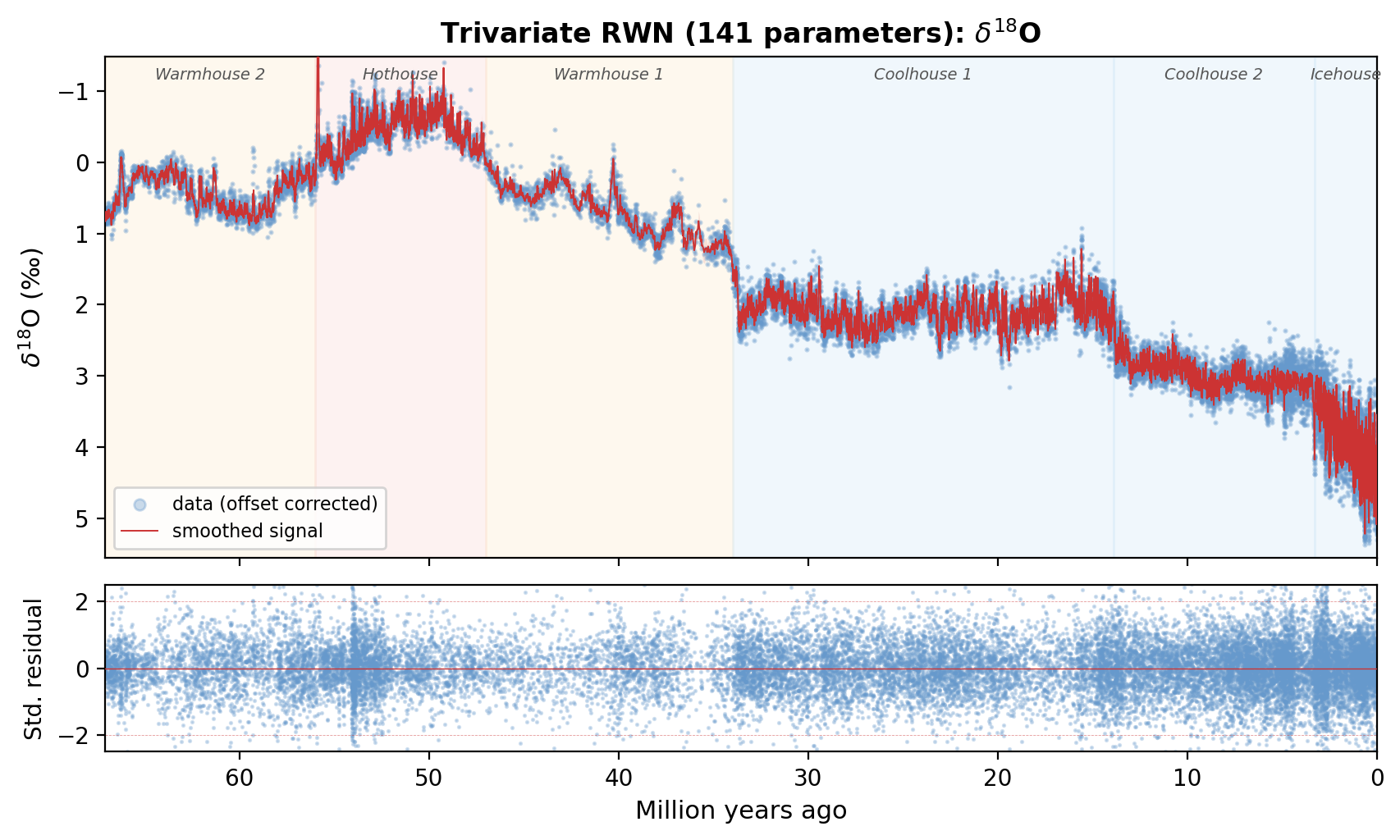}
\caption{\footnotesize Trivariate RWN (141~parameters): $\delta^{18}$O. Top panel: data with their estimated site offsets removed (blue dots) and smoothed state (red line) with 95\% confidence band. Note the inverted $y$-axis. Bottom panel: standardized prediction residuals.\label{F:Milank_rwn_d18O}}
\end{figure}

\begin{figure}[htbp]
\centering
\includegraphics[width=\textwidth]{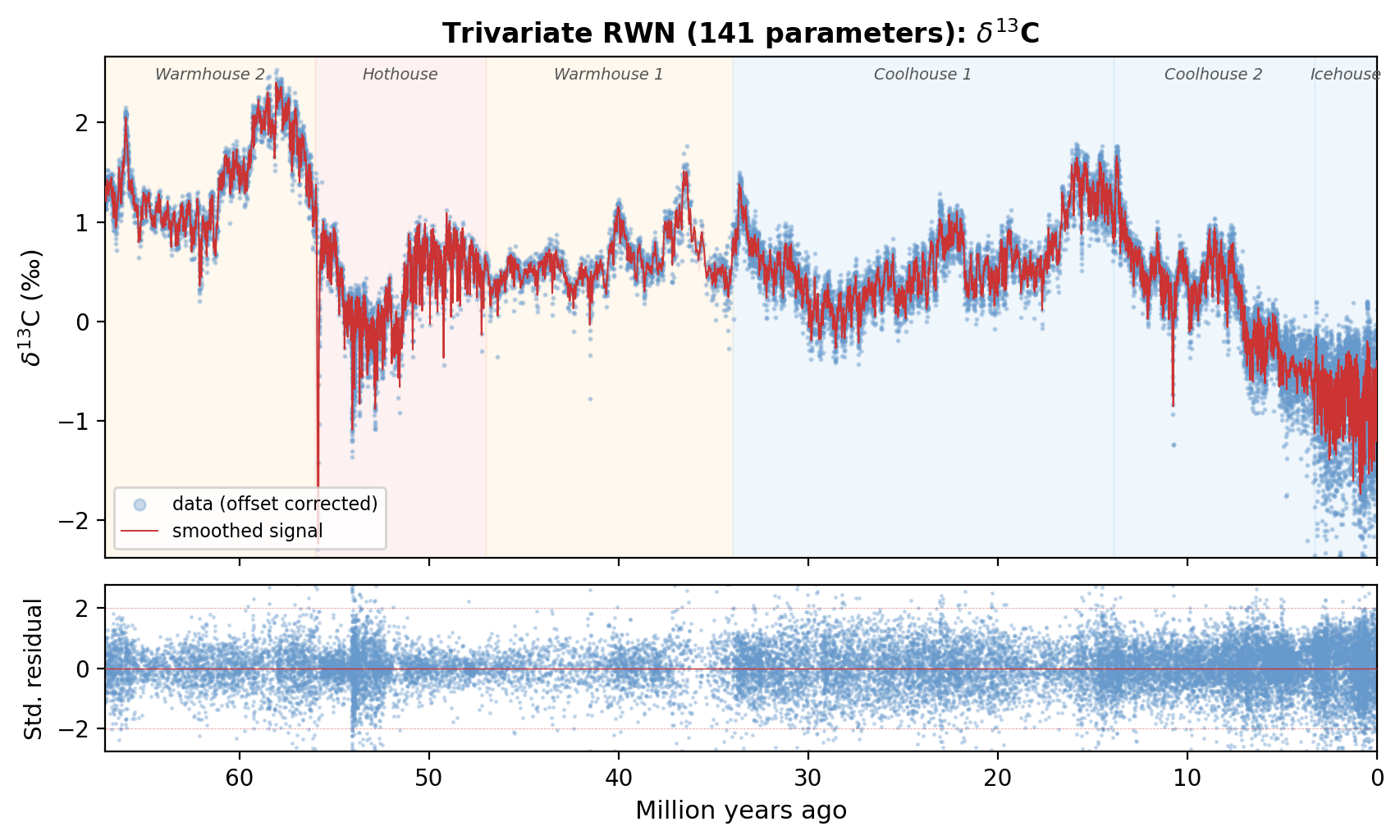}
\caption{\footnotesize Trivariate RWN (141~parameters): $\delta^{13}$C. Top panel: data with their estimated site offsets removed (blue dots) and smoothed state (red line) with 95\% confidence band. Bottom panel: standardized prediction residuals.\label{F:Milank_rwn_d13C}}
\end{figure}

\begin{figure}[htbp]
\centering
\includegraphics[width=\textwidth]{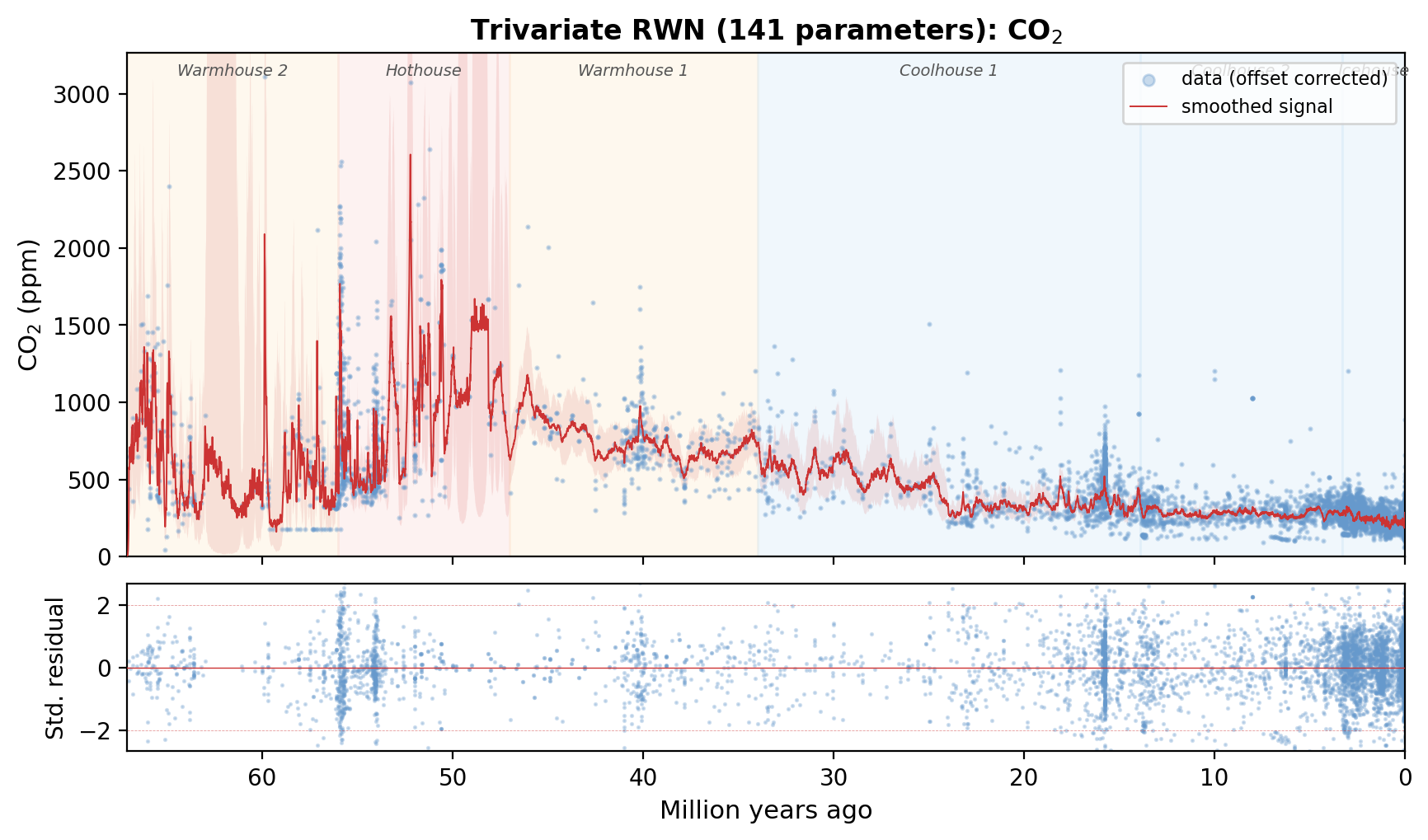}
\caption{\footnotesize Trivariate RWN (141~parameters): CO$_2$. Top panel: CO$_2$ observations with their estimated group offsets removed (blue dots) and smoothed state (red line) with 95\% confidence band, displayed in ppm. Bottom panel: standardized prediction residuals.\label{F:Milank_rwn_CO2}}
\end{figure}

Figures~\ref{F:Milank_rwn_d18O}--\ref{F:Milank_rwn_CO2} show the data and smoothed states from the preferred period-dependent RWN + Milankovitch model. The orbital forcing, while statistically decisive, enters as small increments that mostly average out over the smoothing window, so it sharpens the latent path locally without materially shifting it. The standardized residuals remain well behaved.

A notable feature is the behavior across the long CO$_2$ data gaps, where the CO$_2$ record is silent for up to $\sim$0.9~Myr (most prominently in the Hothouse near $48$--$50$~Ma). There the trivariate random walk reconstructs a bounded, physically plausible CO$_2$ path that joins continuously to the flanking observations, drawing on the densely sampled isotopes through the estimated cross-proxy correlations (which shift the gap-interior CO$_2$ level by up to about $11\%$ relative to a CO$_2$-only reconstruction) and, to a lesser degree, on the deterministic orbital forcing in the preferred model (up to about $3\%$). This contrasts sharply with the integrated random walk (Section~\ref{S:TriIWN2Results}), whose slope state extrapolates an implausible excursion across the same gap (Figure~\ref{F:Milank_iwn2_CO2}), and it is a further argument for the random-walk specification: it bridges the gaps in the sparse CO$_2$ record with information from the other proxies rather than with an unconstrained extrapolation.

\subsection{Comparison with the bivariate model}\label{S:TriVsBiv}

The bias-corrected trivariate model (141~parameters, with period-dependent Milankovitch forcing) achieves a log-likelihood of $21{,}175$. For comparison, the bias-corrected bivariate site-based model of \cite{bhkl2024} achieves a log-likelihood of $21{,}509$ for $\delta^{18}$O and $\delta^{13}$C jointly; the two values are not directly comparable, since the trivariate log-likelihood also includes the 6{,}300 $\log(\text{CO}_2)$ observations, so the meaningful comparison is of the isotope signal extraction itself. Both models place the isotope proxies on a common latent scale through site-specific intercepts and site-specific measurement variances, making the isotope components directly comparable. The $\delta^{18}$O--$\delta^{13}$C correlations in Table~\ref{T:MilankCorr} are very close to those in the bivariate model ($+0.81$ vs.\ $+0.81$ for Warmhouse~2, $+0.97$ vs.\ $+0.97$ for Hothouse, $-0.77$ vs.\ $-0.79$ for Icehouse), confirming that the addition of the CO$_2$ dimension does not distort the isotope signal extraction.

\subsection{Trivariate IWN(2)}\label{S:TriIWN2Results}

The integrated-random-walk counterpart of the preferred model is the trivariate IWN(2) smoothing model of Section~\ref{S:TriIWN2}, with its per-period slope-innovation variances fixed at the univariate estimates ($69$ freely estimated parameters, or $78$ once the nine constant orbital coefficients are added). Its heavier smoothing is a deliberate design choice rather than a deficiency: the IWN(2) is retained not as a competitor on goodness-of-fit but as a signal-extraction device whose smoothed paths attenuate high-frequency noise. Within the IWN(2), adding the orbital block improves the BIC by 280~units ($\text{LR}=377.8$ on 9~degrees of freedom). A direct BIC comparison with the random walk is in any case not a clean nested one, since the 18 slope-innovation variances of the IWN(2) are fixed rather than freely estimated.

\begin{figure}[htbp]
\centering
\includegraphics[width=\textwidth]{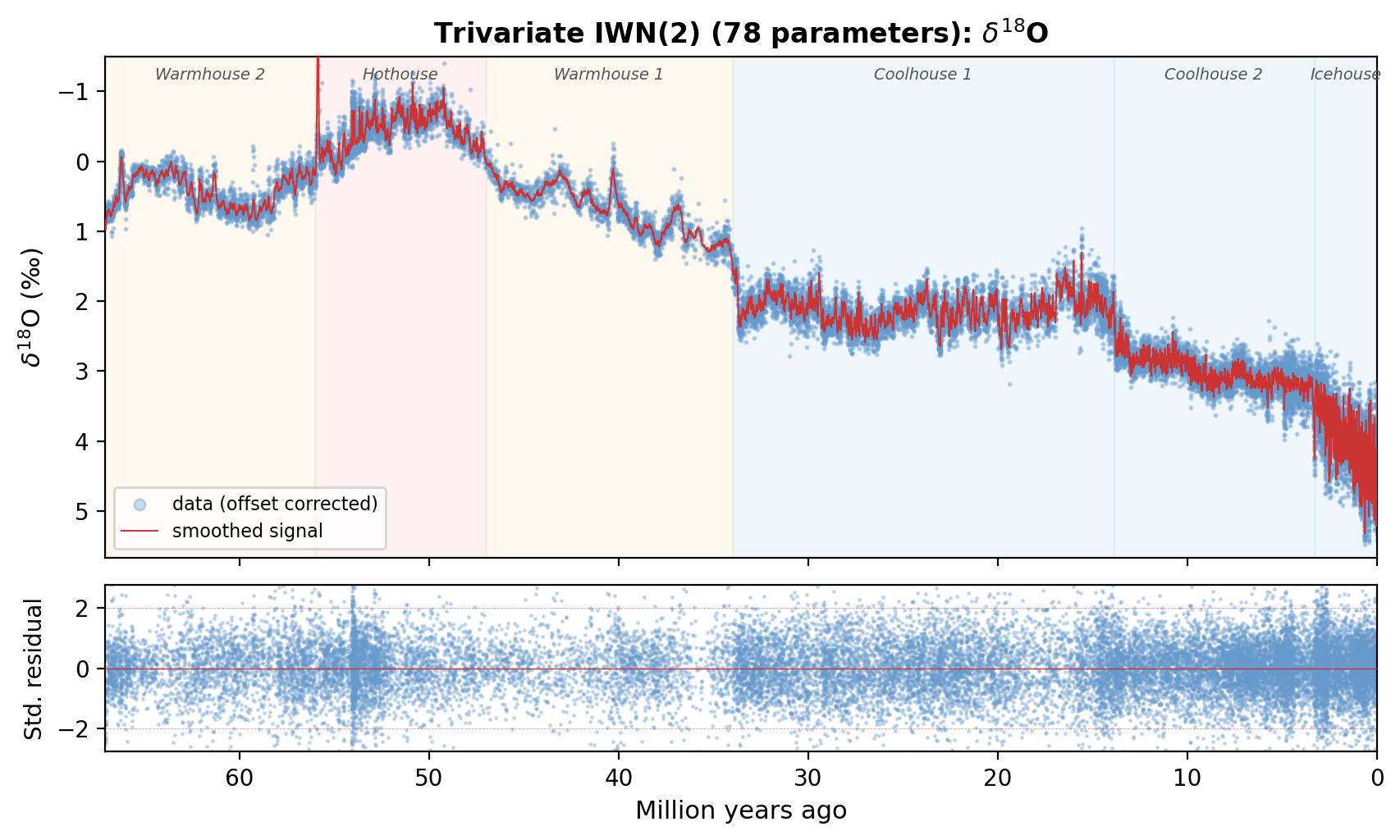}
\caption{\footnotesize Trivariate IWN(2) (78~parameters): $\delta^{18}$O. Top panel: data with their estimated site offsets removed (blue dots) and smoothed level state (red line) with 95\% confidence band. Note the inverted $y$-axis. Bottom panel: standardized prediction residuals.\label{F:Milank_iwn2_d18O}}
\end{figure}

\begin{figure}[htbp]
\centering
\includegraphics[width=\textwidth]{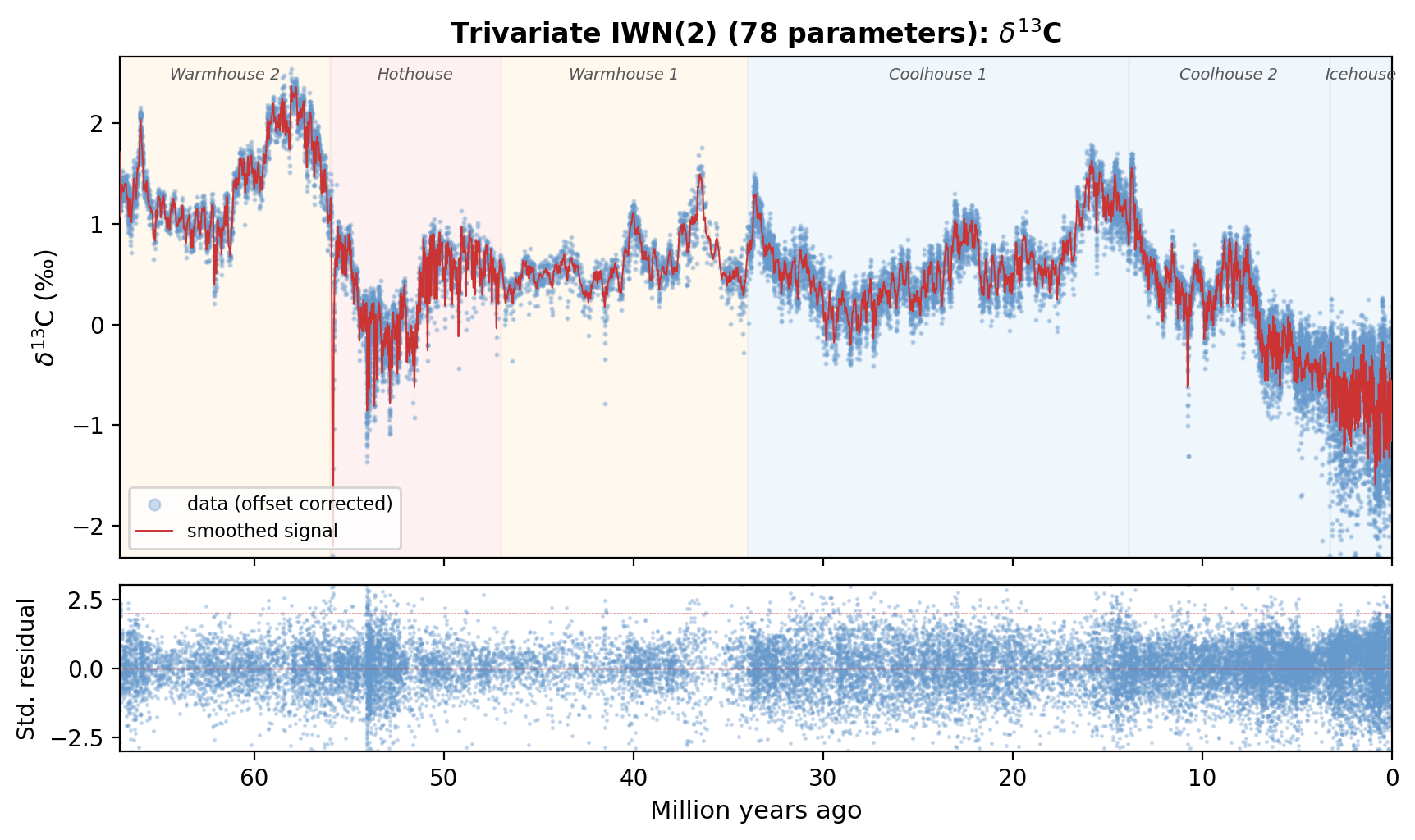}
\caption{\footnotesize Trivariate IWN(2) (78~parameters): $\delta^{13}$C. Top panel: data with their estimated site offsets removed (blue dots) and smoothed level state (red line) with 95\% confidence band. Bottom panel: standardized prediction residuals.\label{F:Milank_iwn2_d13C}}
\end{figure}

\begin{figure}[htbp]
\centering
\includegraphics[width=\textwidth]{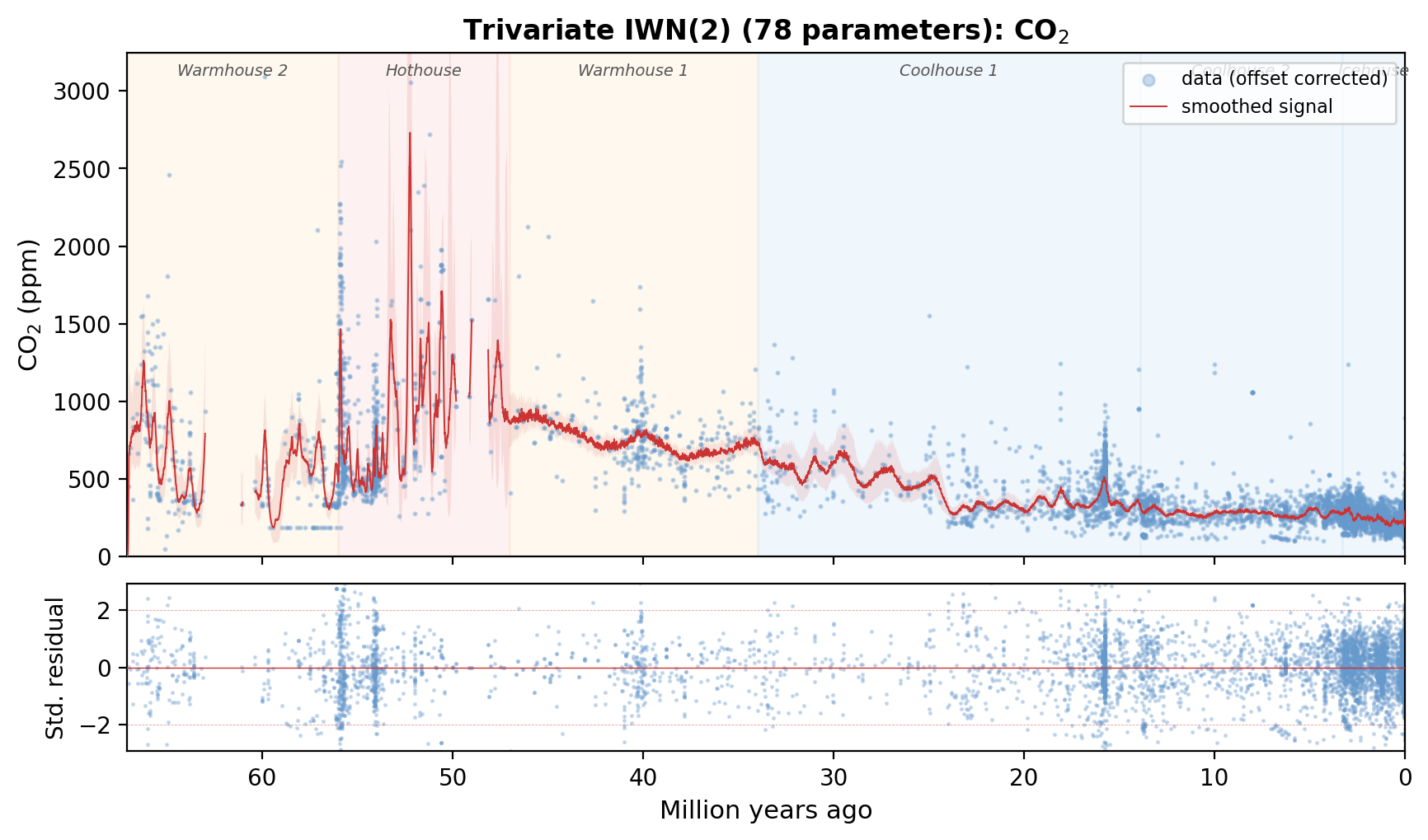}
\caption{\footnotesize Trivariate IWN(2) (78~parameters): CO$_2$. Top panel: CO$_2$ observations with their estimated group offsets removed (blue dots) and smoothed level state (red line) with 95\% confidence band, displayed in ppm. The smoothed level and band are omitted across CO$_2$ data gaps longer than $0.5$~Myr, where the integrated random walk extrapolates without CO$_2$ observations to constrain it. Bottom panel: standardized prediction residuals.\label{F:Milank_iwn2_CO2}}
\end{figure}

Figures~\ref{F:Milank_iwn2_d18O}--\ref{F:Milank_iwn2_CO2} show the data and smoothed states from the trivariate IWN(2) model. The smoothed signals reproduce the main features of the RWN fits (Figures~\ref{F:Milank_rwn_d18O}--\ref{F:Milank_rwn_CO2}) but are visibly smoother, attenuating the sharpest excursions, most clearly the transient CO$_2$ spikes in the Eocene and the short-lived Icehouse excursions, exactly as in the univariate comparison of Figures~\ref{F:CO2_uni} and~\ref{F:CO2_iwn2}. Fixing the slope variances removes the divergence of the free-slope specification, and the CO$_2$ path is well behaved wherever the record is reasonably sampled. Across the few CO$_2$ gaps longer than $0.5$~Myr, most notably in the Hothouse near $48$--$50$~Ma, the order-two integrated random walk nonetheless extrapolates a weakly constrained excursion with a rapidly widening band, its smoothed level variance growing as the cube of the gap length against the linear growth of the random walk; we therefore omit the smoothed CO$_2$ level across these gaps in Figure~\ref{F:Milank_iwn2_CO2}.

\subsubsection*{Orbital forcing coefficients}

Table~\ref{T:MilankOrb_IWN2} reports the estimated orbital coefficients, here held constant across climate states. Their pattern echoes the climate-state-averaged response of the period-dependent RWN model (Table~\ref{T:MilankOrb}): \emph{eccentricity} exerts the strongest and most broadly significant influence, with higher eccentricity associated with lower $\delta^{18}$O (warming and ice loss) and $\delta^{13}$C but higher $\log\text{CO}_2$; \emph{obliquity} acts strongly on $\delta^{18}$O ($t = -9.3$) but is insignificant for the other two proxies; and \emph{climatic precession} has small but significant effects on the isotopes and an insignificant effect on $\log\text{CO}_2$, again plausibly reflecting the limited temporal resolution of the CO$_2$ proxy data at the orbital bands.

\begin{table}[htbp]
\caption{\footnotesize Estimated orbital forcing coefficients $\hat{b}_{ij}$ from the trivariate IWN(2) + Milank model (78~parameters). Standard errors in parentheses, computed by inverting the numerical Hessian. Significance: ${}^{***}p<0.01$, ${}^{**}p<0.05$.\label{T:MilankOrb_IWN2}}
\centering
\begin{tabular}{lrrr}\hline
 & $\delta^{18}$O & $\delta^{13}$C & $\log\text{CO}_2$ \\\hline
Eccentricity       & $-1.213^{***}$ & $-2.194^{***}$ & $+1.223^{***}$ \\[4pt]
              & $(0.143)$ & $(0.140)$ & $(0.421)$ \\
Obliquity          & $-1.177^{***}$ & $+0.155$ & $+0.805$ \\[4pt]
              & $(0.126)$ & $(0.119)$ & $(0.425)$ \\
Clim.\ precession  & $+0.192^{***}$ & $-0.094^{**}$ & $-0.218$ \\
              & $(0.041)$ & $(0.039)$ & $(0.164)$ \\
\hline
\end{tabular}
\end{table}

\subsubsection*{Measurement noise by proxy group}

Table~\ref{T:TriIWN2_eps} reports the estimated CO$_2$ proxy-group intercepts $\hat c_g$ and measurement standard deviations $\hat\sigma_{\eps,g}$. The intercepts place all proxies on the common \texttt{b\_ca} latent scale: the boron isotopes sit about $0.14$ in $\log(\text{CO}_2)$ above the reference and the small nahcolite group about $0.81$ above it, while land-plant $\delta^{13}$C, paleosol, phytoplankton, and the pooled stomata group sit within $\pm 0.07$ of the reference. Conditional on these offsets, the residual measurement noise varies substantially across proxy groups: nahcolite and paleosol are the noisiest ($\hat\sigma_\eps \approx 0.60$--$0.70$), the boron isotopes and land-plant $\delta^{13}$C the most precise ($\hat\sigma_\eps \approx 0.23$), with phytoplankton and pooled stomata intermediate ($\approx 0.34$--$0.36$). These values are close to those from the univariate RWN and IWN(2) of Table~\ref{T:CO2_eps}, the small differences reflecting the trivariate coupling. (The drill-site intercepts and measurement variances for $\delta^{18}$O and $\delta^{13}$C are as in the bivariate model of \cite{bhkl2024} and are not reproduced here.)

\begin{table}[htbp]
\caption{\footnotesize Estimated CO$_2$ proxy-group intercepts $\hat c_g$ and measurement standard deviations $\hat\sigma_{\eps,g}$ from the trivariate IWN(2) model. Measurement variances for CO$_2$ are differentiated by proxy group, pooling the six stomatal sub-proxies and liverwort into a single \emph{stomata} group; \texttt{b\_ca} is the intercept reference ($c_g=0$). Standard errors in parentheses for both the intercept and the measurement standard deviation (the latter by the delta method from the Hessian-based standard error of the estimated variance).\label{T:TriIWN2_eps}}
\centering
\begin{tabular}{lrrr}\hline
CO$_2$ proxy group & No.\ obs. & $\hat c_g$ & $\hat\sigma_{\eps,g}$ \\\hline
\texttt{b\_ca}           & 277 & $0$ (ref.) & 0.159 $(0.009)$ \\
Boron isotopes           & 2{,}111 & $+0.141$ $(0.013)$ & 0.235 $(0.004)$ \\
Land plant $\delta^{13}$C & 1{,}799 & $-0.047$ $(0.014)$ & 0.231 $(0.004)$ \\
Nahcolite                & 4 & $+0.811$ $(0.348)$ & 0.698 $(0.303)$ \\
Paleosol                 & 319 & $-0.052$ $(0.037)$ & 0.588 $(0.026)$ \\
Phytoplankton            & 1{,}201 & $+0.029$ $(0.016)$ & 0.356 $(0.008)$ \\
Stomata (pooled)         & 589 & $-0.070$ $(0.022)$ & 0.337 $(0.012)$ \\
\hline
\end{tabular}
\end{table}

\subsubsection*{Slope-innovation correlations}

Table~\ref{T:MilankCorr_IWN2} reports the pairwise slope-innovation correlations of the IWN(2) + Milankovitch model by climate state. They track the RWN level correlations (Table~\ref{T:MilankCorr}) and the bivariate analysis of \cite{bhkl2024}: the $\delta^{18}$O--$\delta^{13}$C correlation is strongly positive in the warm and early-cool periods and reverses to strongly negative in the Coolhouse~2 and Icehouse, while the $\delta^{18}$O--$\log\text{CO}_2$ correlation is predominantly negative, confirming that the smoothing model recovers the same cross-proxy covariation structure.

\begin{table}[htbp]
\caption{\footnotesize Maximum likelihood estimates of pairwise slope-innovation correlations $\hat\rho_k$ from the trivariate IWN(2) + Milank model (78~parameters), by climate state. 95\% confidence intervals (in brackets) are obtained from the Hessian-based standard errors via the Fisher-$z$ transform; the $\delta^{18}$O--CO$_2$ and $\delta^{13}$C--CO$_2$ correlations are weakly identified in several states, hence the wide intervals. Climate states as defined in \cite{westerhold2020}.\label{T:MilankCorr_IWN2}}
\centering\footnotesize
\begin{tabular}{lrrr}\hline
Climate state & $\hat\rho_{12,k}$ & $\hat\rho_{13,k}$ & $\hat\rho_{23,k}$ \\
 & ($\delta^{18}$O, $\delta^{13}$C) & ($\delta^{18}$O, $\log\text{CO}_2$) & ($\delta^{13}$C, $\log\text{CO}_2$) \\\hline
Warmhouse 2 (67--56 Ma)    & $+0.871$ & $+0.187$ & $+0.438$ \\
                           & $[+0.82,\,+0.91]$ & $[-0.64,\,+0.81]$ & $[-0.51,\,+0.90]$ \\[3pt]
Hothouse (56--47 Ma)       & $+0.989$ & $-0.466$ & $-0.456$ \\
                           & $[+0.98,\,+0.99]$ & $[-0.70,\,-0.15]$ & $[-0.70,\,-0.12]$ \\[3pt]
Warmhouse 1 (47--34 Ma)    & $+0.756$ & $-0.367$ & $+0.189$ \\
                           & $[+0.63,\,+0.84]$ & $[-0.92,\,+0.69]$ & $[-0.76,\,+0.88]$ \\[3pt]
Coolhouse 1 (34--13.9 Ma)  & $+0.955$ & $+0.092$ & $+0.216$ \\
                           & $[+0.94,\,+0.97]$ & $[-0.64,\,+0.74]$ & $[-0.55,\,+0.78]$ \\[3pt]
Coolhouse 2 (13.9--3.3 Ma) & $-0.505$ & $-0.130$ & $-0.282$ \\
                           & $[-0.65,\,-0.32]$ & $[-0.99,\,+0.99]$ & $[-0.95,\,+0.85]$ \\[3pt]
Icehouse (3.3--0 Ma)       & $-0.991$ & $-0.530$ & $+0.414$ \\
                           & $[-0.99,\,-0.98]$ & $[-0.77,\,-0.17]$ & $[+0.03,\,+0.69]$ \\
\hline
\end{tabular}
\end{table}

\subsubsection*{Fixed state variance rates}

Table~\ref{T:TriEta_IWN2} reports the per-period slope-innovation standard deviations used in the trivariate IWN(2) model. These are not estimated jointly but fixed at the univariate values, as described in Section~\ref{S:TriIWN2}. As in the univariate case (Table~\ref{T:CO2_eta}), the slope-innovation rates are numerically large because they enter the level only after a double time integration, carrying a $\Delta t^{-(2m-1)}$ scaling; they are not directly comparable to the RWN level-innovation rates of Table~\ref{T:MilankEta}.

\begin{table}[htbp]
\caption{\footnotesize Per-period slope-innovation standard deviations $\hat\sigma_{\eta,k}$ used in the trivariate IWN(2) model, by climate state and proxy. These are \emph{fixed} at the estimates from the corresponding univariate pure $m$-fold IWN(2) specifications (drill-site / proxy-type measurement variances with group intercepts and per-period state variances), not re-estimated jointly. Standard errors in parentheses are from those univariate fits (delta method), where the variances are estimated. As in the univariate case (Table~\ref{T:CO2_eta}), the slope-innovation rates are large because they enter the level only after a double time integration ($\Delta t^{-(2m-1)}$ scaling).\label{T:TriEta_IWN2}}
\centering\footnotesize
\begin{tabular}{lrrr}\hline
Climate state & $\hat\sigma_{\eta,k}^{\delta^{18}O}$ & $\hat\sigma_{\eta,k}^{\delta^{13}C}$ & $\hat\sigma_{\eta,k}^{\log\text{CO}_2}$ \\\hline
Warmhouse 2 (67--56 Ma)    & 14.17 & 38.09 & 10.01 \\
                           & $(1.45)$ & $(2.22)$ & $(2.81)$ \\[3pt]
Hothouse (56--47 Ma)       & 75.05 & 146.19 & 38.04 \\
                           & $(4.82)$ & $(7.33)$ & $(5.34)$ \\[3pt]
Warmhouse 1 (47--34 Ma)    & 3.58 & 6.21 & 0.17 \\
                           & $(0.44)$ & $(0.78)$ & $(0.08)$ \\[3pt]
Coolhouse 1 (34--13.9 Ma)  & 73.45 & 30.83 & 1.95 \\
                           & $(4.27)$ & $(1.94)$ & $(0.40)$ \\[3pt]
Coolhouse 2 (13.9--3.3 Ma) & 163.91 & 142.03 & 1.11 \\
                           & $(8.76)$ & $(8.17)$ & $(0.31)$ \\[3pt]
Icehouse (3.3--0 Ma)       & 735.06 & 444.64 & 23.47 \\
                           & $(34.95)$ & $(33.66)$ & $(3.44)$ \\
\hline
\end{tabular}
\end{table}

\clearpage

\section{Smoothed CO$_2$ across major climate transitions}\label{S:Transitions}

\subsection{Smoothed CO$_2$ across the Eocene--Oligocene transition}\label{S:MilankEOT}

Table~\ref{T:EOT_ext} reports the smoothed CO$_2$ across an extended window (38--32~Ma) from four model specifications: the two trivariate models (the preferred RWN, with and without Milankovitch forcing) and the two univariate CO$_2$ models (RWN and IWN(2)). The trivariate and univariate estimates are very close at every age, confirming that the CO$_2$ reconstruction is robust to both the choice of stochastic specification and the inclusion of orbital forcing or isotope data.

Each of Tables~\ref{T:EOT_ext}, \ref{T:Miocene}, and~\ref{T:Greenland} also reports a simple model-free benchmark in its final column: a centered moving average of the offset-corrected CO$_2$ proxies, with the averaging half-window chosen by $h$-block cross-validation within each transition window (0.1--0.2~Myr). The block deletion is essential here: deleting only the test point, as in ordinary cross-validation, lets near-coincident replicate measurements from the same study predict one another and drives the selected window to a degenerate near-zero width; removing a small neighbourhood around each test point instead yields a stable bandwidth. The moving average recovers essentially the same central CO$_2$ path as the state-space smoothers. We report it as a \emph{point estimate only}, and this is deliberate. The natural standard error of a window mean, $s/\sqrt{n}$, presumes that the proxies inside the window are independent draws about a locally constant level; they are neither, being serially dependent and heterogeneously precise samples of a non-stationary (integrated) signal that is itself changing across the window, so $s/\sqrt{n}$ would badly understate the true uncertainty. Furnishing \emph{calibrated} confidence intervals, rather than a deceptively tight one, is precisely an advantage of the state-space approach, and it is why we attach uncertainty bands only to the model-based columns.

\begin{table}[htbp]
\caption{\footnotesize Smoothed CO$_2$ (ppm) across the Eocene--Oligocene transition (38--32~Ma) from the two trivariate and two univariate CO$_2$ model specifications.  The trivariate RWN\,+\,Milank column uses the preferred period-dependent Milankovitch model.  The 95\% confidence intervals are from the Kalman smoother.  The final column (MA) is a centered moving average of the offset-corrected CO$_2$ proxies over a $\pm0.2$~Myr window selected by $h$-block cross-validation, reported as a point estimate only (see text).  The Warmhouse~1/Coolhouse~1 boundary is at 34~Ma (bold).\label{T:EOT_ext}}
\centering
\setlength{\tabcolsep}{4pt}\footnotesize
\begin{tabular}{r cl cl cl cl r}
\toprule
 & \multicolumn{4}{c}{Trivariate} & \multicolumn{4}{c}{Univariate CO$_2$} & \\
\cmidrule(lr){2-5} \cmidrule(lr){6-9}
 & \multicolumn{2}{c}{RWN} & \multicolumn{2}{c}{RWN + Milank} & \multicolumn{2}{c}{RWN} & \multicolumn{2}{c}{IWN(2)} & MA \\
\cmidrule(lr){2-3} \cmidrule(lr){4-5} \cmidrule(lr){6-7} \cmidrule(lr){8-9} \cmidrule(lr){10-10}
Age (Ma) & ppm & 95\,\% CI & ppm & 95\,\% CI & ppm & 95\,\% CI & ppm & 95\,\% CI & ppm \\
\midrule
38.0 & 577 & [490, 680] & 575 & [485, 681] & 577 & [488, 683] & 638 & [582, 699] & 531 \\
37.5 & 632 & [510, 785] & 628 & [500, 787] & 620 & [495, 776] & 631 & [574, 695] & 677 \\
37.0 & 711 & [600, 843] & 700 & [587, 836] & 705 & [587, 847] & 648 & [588, 714] & 737 \\
36.5 & 653 & [527, 809] & 630 & [503, 789] & 630 & [505, 785] & 663 & [600, 732] & 587 \\
36.0 & 662 & [551, 795] & 668 & [552, 807] & 658 & [545, 794] & 679 & [615, 750] & 654 \\
35.5 & 653 & [546, 780] & 658 & [547, 792] & 649 & [540, 778] & 697 & [629, 772] & 652 \\
35.0 & 673 & [528, 858] & 677 & [524, 873] & 681 & [529, 877] & 718 & [646, 798] & 709 \\
34.5 & 757 & [629, 909] & 757 & [624, 918] & 748 & [623, 898] & 735 & [665, 813] & 733 \\
\textbf{34.0} & \textbf{737} & \textbf{[602, 901]} & \textbf{738} & \textbf{[601, 907]} & \textbf{721} & \textbf{[586, 885]} & \textbf{734} & \textbf{[644, 837]} & \textbf{681} \\
33.5 & 593 & [474, 741] & 590 & [473, 735] & 591 & [481, 727] & 606 & [535, 686] & 555 \\
33.0 & 580 & [462, 729] & 577 & [460, 723] & 581 & [484, 698] & 600 & [515, 698] & 620 \\
32.5 & 562 & [398, 795] & 558 & [396, 787] & 561 & [403, 781] & 579 & [455, 737] & 512 \\
32.0 & 579 & [433, 774] & 587 & [440, 784] & 598 & [448, 799] & 544 & [427, 694] & 664 \\
\bottomrule
\end{tabular}
\end{table}

All four model specifications give broadly similar estimates. Smoothed CO$_2$ at 34~Ma is 721--738~ppm across the models. A local hump at $\sim$34.5~Ma ($\sim$735--757~ppm) is followed by a sustained decline into the Oligocene, reaching $\sim$544--598~ppm by 32~Ma, a reduction of approximately 20\%. This CO$_2$ drawdown is consistent with the widely discussed role of declining atmospheric CO$_2$ as one contributing driver, alongside ocean-gateway and circulation changes, of the onset of large-scale Antarctic glaciation at the Eocene--Oligocene boundary \citep{zachos2001}; see \citet{goldner2014}, \citet{hutchinson2019}, and \citet{straume2022} for the causes of the transition.

To keep Table~\ref{T:EOT_ext} compact we omit the trivariate IWN(2) smoothing model (Section~\ref{S:TriIWN2Results}); its smoothed CO$_2$ at the 34~Ma boundary is 735~ppm (95\% CI [651, 830]) without orbital forcing and 734~ppm ([649, 831]) with Milankovitch forcing, slightly above the random-walk estimates but consistent with the univariate IWN(2) column (734~ppm) and reflecting the same mild upward shift from heavier smoothing. The IWN(2) path traces the same hump-and-decline, peaking near 34.5~Ma (726~ppm [662, 797] without forcing; 736~ppm [670, 809] with) and falling to 548~ppm ([433, 694]) and 550~ppm ([434, 698]), respectively, by 32~Ma.

For comparison, \cite{hansen2008target} inferred CO$_2 \approx 450$~ppm at 35~Ma by calibrating a fast-feedback climate model to deep-ocean $\delta^{18}$O data. Their estimate is lower than all of our smoothed values at the same age (673--718~ppm). Because \cite{hansen2008target} did not employ a formal statistical model, they could not report confidence intervals; the difference may partly reflect the additional proxy data now available through the CenCO2PIP compilation \citep{cenco2pip2023}, as well as the distinct identification strategies.

The close agreement between the trivariate and univariate RWN smoothed values indicates that the isotope data, which are much denser than the CO$_2$ data in this interval, do not substantially pull the CO$_2$ state estimate away from its univariate value, consistent with the weak correlations between isotopes and $\log\text{CO}_2$ estimated for the Warmhouse~1 and Coolhouse~1 periods (Table~\ref{T:MilankCorr}).

\subsection{Smoothed CO$_2$ across the Miocene glaciation of West Antarctica}\label{S:MioceneGlac}

The middle Miocene ($\sim$14~Ma) marks a major intensification of Antarctic glaciation and the transition from Coolhouse~1 to Coolhouse~2 at 13.9~Ma. Table~\ref{T:Miocene} reports the smoothed CO$_2$ from the two trivariate and two univariate model specifications across a window spanning 16--11~Ma. As with the EOT, the estimates are broadly consistent across models.

The trivariate IWN(2) smoothing model (Section~\ref{S:TriIWN2Results}), again omitted from Table~\ref{T:Miocene} for space, gives smoothed CO$_2$ of 324~ppm (95\% CI [302, 348]) at the 13.9~Ma boundary without orbital forcing and 328~ppm ([306, 352]) with Milankovitch forcing, essentially identical to the univariate IWN(2) (328~ppm) and within the spread of the tabulated models. From $\sim$409~ppm ([385, 435]) at 16~Ma it declines to 265~ppm ([232, 302]) and 260~ppm ([228, 297]), respectively, by 11~Ma.

\begin{table}[htbp]
\caption{\footnotesize Smoothed CO$_2$ (ppm) across the Miocene West Antarctica glaciation (16--11~Ma) from the two trivariate and two univariate CO$_2$ model specifications.  The trivariate RWN\,+\,Milank column uses the preferred period-dependent Milankovitch model.  The final column (MA) is a centered moving average of the offset-corrected CO$_2$ proxies over a $\pm0.2$~Myr window selected by $h$-block cross-validation, reported as a point estimate only (see text).  The Coolhouse~1/Coolhouse~2 boundary is at 13.9~Ma (bold).\label{T:Miocene}}
\centering
\setlength{\tabcolsep}{4pt}\footnotesize
\begin{tabular}{r cl cl cl cl r}
\toprule
 & \multicolumn{4}{c}{Trivariate} & \multicolumn{4}{c}{Univariate CO$_2$} & \\
\cmidrule(lr){2-5} \cmidrule(lr){6-9}
 & \multicolumn{2}{c}{RWN} & \multicolumn{2}{c}{RWN + Milank} & \multicolumn{2}{c}{RWN} & \multicolumn{2}{c}{IWN(2)} & MA \\
\cmidrule(lr){2-3} \cmidrule(lr){4-5} \cmidrule(lr){6-7} \cmidrule(lr){8-9} \cmidrule(lr){10-10}
Age (Ma) & ppm & 95\,\% CI & ppm & 95\,\% CI & ppm & 95\,\% CI & ppm & 95\,\% CI & ppm \\
\midrule
16.0 & 379 & [332, 433] & 372 & [326, 425] & 383 & [336, 436] & 408 & [383, 435] & 413 \\
15.5 & 307 & [264, 357] & 311 & [267, 361] & 312 & [269, 362] & 346 & [316, 380] & 349 \\
15.0 & 367 & [322, 419] & 367 & [322, 418] & 371 & [326, 422] & 350 & [323, 378] & 355 \\
14.5 & 305 & [250, 371] & 303 & [249, 369] & 308 & [253, 375] & 298 & [263, 338] & 294 \\
14.0 & 434 & [369, 510] & 434 & [369, 509] & 436 & [371, 513] & 355 & [327, 387] & 385 \\
\textbf{13.9} & \textbf{338} & \textbf{[302, 378]} & \textbf{340} & \textbf{[304, 380]} & \textbf{338} & \textbf{[301, 380]} & \textbf{328} & \textbf{[305, 352]} & \textbf{303} \\
13.5 & 323 & [296, 353] & 320 & [293, 350] & 318 & [290, 349] & 302 & [283, 321] & 282 \\
13.0 & 319 & [290, 352] & 320 & [291, 353] & 325 & [294, 359] & 315 & [292, 341] & 319 \\
12.5 & 267 & [235, 304] & 265 & [233, 302] & 271 & [237, 311] & 268 & [243, 296] & 255 \\
12.0 & 304 & [269, 344] & 308 & [273, 349] & 299 & [263, 340] & 292 & [264, 324] & 264 \\
11.5 & 269 & [233, 311] & 265 & [229, 306] & 273 & [236, 317] & 282 & [252, 316] & 272 \\
11.0 & 273 & [233, 319] & 275 & [235, 322] & 278 & [236, 329] & 265 & [232, 304] & 259 \\
\bottomrule
\end{tabular}
\end{table}

\subsection{Smoothed CO$_2$ across the Greenland glaciation}\label{S:GreenlandGlac}

The late Pliocene--early Pleistocene ($\sim$3.3--2.6~Ma) saw the onset of Northern Hemisphere glaciation, including the Greenland ice sheet, and marks the transition from Coolhouse~2 to the Icehouse at 3.3~Ma. Table~\ref{T:Greenland} reports the smoothed CO$_2$ from the four model specifications in steps of 200~kyr across this transition. Once again, the estimates are broadly robust across all model specifications.

The trivariate IWN(2) smoothing model (Section~\ref{S:TriIWN2Results}), not tabulated here for space, places smoothed CO$_2$ at the 3.3~Ma Coolhouse~2/Icehouse boundary at 276~ppm (95\% CI [260, 292]) without orbital forcing and 272~ppm ([256, 289]) with Milankovitch forcing, consistent with the univariate IWN(2) (273~ppm) and the other specifications. Across the window it falls from 272~ppm ([257, 287]) and 265~ppm ([251, 280]) at 4.0~Ma to 247~ppm ([225, 270]) and 243~ppm ([222, 265]) at 2.0~Ma.

\begin{table}[htbp]
\caption{\footnotesize Smoothed CO$_2$ (ppm) across the Greenland glaciation and onset of Northern Hemisphere ice sheets (4.0--2.0~Ma) from the two trivariate and two univariate CO$_2$ model specifications.  The trivariate RWN\,+\,Milank column uses the preferred period-dependent Milankovitch model.  The final column (MA) is a centered moving average of the offset-corrected CO$_2$ proxies over a $\pm0.1$~Myr window selected by $h$-block cross-validation, reported as a point estimate only (see text).  The Coolhouse~2/Icehouse boundary is at 3.3~Ma (bold).  Steps of 200~kyr.\label{T:Greenland}}
\centering
\setlength{\tabcolsep}{4pt}\footnotesize
\begin{tabular}{r cl cl cl cl r}
\toprule
 & \multicolumn{4}{c}{Trivariate} & \multicolumn{4}{c}{Univariate CO$_2$} & \\
\cmidrule(lr){2-5} \cmidrule(lr){6-9}
 & \multicolumn{2}{c}{RWN} & \multicolumn{2}{c}{RWN + Milank} & \multicolumn{2}{c}{RWN} & \multicolumn{2}{c}{IWN(2)} & MA \\
\cmidrule(lr){2-3} \cmidrule(lr){4-5} \cmidrule(lr){6-7} \cmidrule(lr){8-9} \cmidrule(lr){10-10}
Age (Ma) & ppm & 95\,\% CI & ppm & 95\,\% CI & ppm & 95\,\% CI & ppm & 95\,\% CI & ppm \\
\midrule
4.0 & 293 & [268, 320] & 291 & [266, 319] & 296 & [270, 324] & 274 & [258, 290] & 323 \\
3.8 & 276 & [249, 306] & 280 & [252, 311] & 276 & [248, 307] & 290 & [271, 310] & 298 \\
3.6 & 290 & [264, 319] & 289 & [263, 318] & 293 & [266, 323] & 290 & [272, 309] & 307 \\
3.4 & 269 & [251, 289] & 271 & [252, 290] & 269 & [251, 288] & 277 & [264, 290] & 282 \\
\textbf{3.3} & \textbf{283} & \textbf{[261, 308]} & \textbf{283} & \textbf{[260, 307]} & \textbf{285} & \textbf{[262, 310]} & \textbf{273} & \textbf{[256, 291]} & \textbf{284} \\
3.2 & 293 & [267, 321] & 291 & [266, 319] & 293 & [265, 325] & 293 & [274, 312] & 277 \\
3.0 & 298 & [271, 327] & 295 & [269, 323] & 298 & [269, 329] & 301 & [278, 326] & 300 \\
2.8 & 268 & [242, 296] & 269 & [243, 297] & 272 & [244, 304] & 277 & [256, 300] & 273 \\
2.6 & 249 & [226, 274] & 249 & [227, 274] & 253 & [228, 281] & 246 & [227, 266] & 249 \\
2.4 & 287 & [259, 318] & 289 & [261, 319] & 289 & [259, 323] & 283 & [261, 307] & 278 \\
2.2 & 249 & [226, 276] & 250 & [226, 275] & 254 & [229, 282] & 251 & [229, 276] & 266 \\
2.0 & 247 & [213, 286] & 247 & [214, 286] & 256 & [217, 303] & 256 & [226, 291] & 252 \\
\bottomrule
\end{tabular}
\end{table}

\subsection{Glaciation thresholds and current CO$_2$ levels}\label{S:Thresholds}

Tables~\ref{T:EOT_ext}, \ref{T:Miocene}, and~\ref{T:Greenland} jointly map the smoothed CO$_2$ across the three major Cenozoic glaciation events. Read in reverse chronological order, they trace successively lower CO$_2$ thresholds associated with the growth of each major ice sheet.

The Greenland glaciation (Table~\ref{T:Greenland}) offers the most direct benchmark for the current climate. Smoothed CO$_2$ in the transition window (4.0--2.0~Ma) ranges from approximately 246 to 301~ppm across all four model specifications, with 95\% confidence intervals reaching at most 329~ppm (at 3.0~Ma). Pre-industrial atmospheric CO$_2$ of approximately 280~ppm lies squarely within these ranges, and current atmospheric CO$_2$ of approximately 425~ppm now exceeds every upper confidence bound at every age in Table~\ref{T:Greenland}, across all four model specifications. These reconstructions date the \emph{onset} of Northern Hemisphere glaciation, so they constrain the CO$_2$ level at which the ice sheet grew, not the level at which an already-established ice sheet would be lost. Because of ice-sheet hysteresis the two need not coincide: ice growth and retreat occur at different CO$_2$ levels \citep{deconto2003}, and the deglaciation threshold, which lies above the glaciation threshold, is not constrained by the record. That present CO$_2$ exceeds the glaciation threshold therefore does not by itself imply that the modern Greenland ice sheet is committed to retreat.

The Miocene glaciation of West Antarctica (Table~\ref{T:Miocene}) marks a higher threshold. Smoothed CO$_2$ around the 13.9~Ma boundary ranges from 328 to 340~ppm across models, with upper confidence bounds reaching 380~ppm. Current CO$_2$ already exceeds the upper confidence bound at every post-transition age from 13.5~Ma onward. At the pre-transition ages (16--14~Ma), where smoothed values reach 298--444~ppm, several upper bounds extend to 440--513~ppm and thus still bracket the current level.

The East Antarctic glaciation at the EOT (Table~\ref{T:EOT_ext}) represents a substantially higher CO$_2$ threshold. Smoothed values at the 34~Ma boundary are 721--738~ppm across models and remain above 544~ppm throughout the 38--32~Ma window. Current CO$_2$ levels are far below these central estimates. However, the lower 95\% confidence bounds at several ages fall into the 400s (for example, 427~ppm at 32.0~Ma and 485~ppm at 38.0~Ma), indicating that the statistical uncertainty does not entirely exclude CO$_2$ levels comparable to those prevailing today.

Taken together, these tables suggest a hierarchy of glaciation thresholds: Northern Hemisphere ice requires CO$_2$ below roughly 280--330~ppm, West Antarctic ice below roughly 330--440~ppm, and East Antarctic ice below roughly 550--760~ppm. Current atmospheric CO$_2$ (about 425~ppm) sits above the Northern Hemisphere and West Antarctic thresholds and below the East Antarctic one (whose lower confidence bounds nonetheless reach into the 400s~ppm). For two of the three glaciations, then, present-day CO$_2$ falls in the range where the persistence of the modern ice sheets turns on hysteresis rather than on these thresholds alone. As noted above, all three records date glaciation onsets and so pin the CO$_2$ levels for ice-sheet \emph{growth}; the higher levels required to remove an already-established ice sheet are not the ones the record constrains.

\section{Conclusion}\label{S:Conclusion}

We have developed and estimated a continuous-time, trivariate state-space model that reconstructs benthic $\delta^{18}$O and $\delta^{13}$C together with atmospheric CO$_2$ over the Cenozoic from irregularly sampled, multi-site, multi-method proxy data, with calibrated uncertainty for the latent signals. The Bayes information criterion prefers the random-walk dynamics over higher-order integrated specifications, and decisively prefers the bias-corrected model with climate-state-dependent innovation covariance and state-dependent Milankovitch forcing.

Three substantive findings emerge. First, the cross-proxy innovation correlation between $\delta^{18}$O and $\delta^{13}$C reverses from strongly positive in the early Cenozoic greenhouse to strongly negative in the icehouse, and the $\delta^{18}$O--CO$_2$ correlation turns negative in the cool states, consistent with rising CO$_2$ accompanying warming and ice loss. Second, the orbital sensitivity of the isotopes is negligible in the ice-free early Cenozoic and strengthens monotonically into the icehouse, mirroring the obliquity pacing of late-Cenozoic glacial cycles; the orbital imprint on CO$_2$ is weak, plausibly reflecting the limited temporal resolution of the CO$_2$ proxy record. Third, the reconstructed CO$_2$ places the thresholds of the three major Cenozoic glaciations in relation to the present: current atmospheric CO$_2$ of about 425~ppm lies above the levels reconstructed around the onset of Northern Hemisphere and West Antarctic glaciation and below those at the Eocene--Oligocene onset of large-scale East Antarctic glaciation. These threshold comparisons should be read within a hysteretic framework, since ice growth and retreat need not occur at the same CO$_2$ level.

The framework extends naturally to incorporate age-model (timestamp) uncertainty, which we treat in ongoing work through an ensemble over astronomically tuned chronologies.


\appendix

\section{System matrices of the state space models}\label{A:Matrices}

The three proxies are merged into a single chronological sequence in long form (Section~\ref{S:TriRWN}): the $N = 54{,}498$ scalar observations are sorted by age, and the Kalman filter performs a scalar measurement update at each step. Denote by $p(n) \in \{1,2,3\}$ the proxy observed at step $n$ ($1 = \delta^{18}$O, $2 = \delta^{13}$C, $3 = \log\text{CO}_2$), by $s(n)$ its drill site (for the isotopes), and by $g(n)$ its proxy group (for CO$_2$).

\subsection{Trivariate random walk plus noise with Milankovitch forcing}

The state vector collects the three latent signals,
\[
\mu_{t_n} = \big(\mu^{\delta^{18}O}_{t_n},\ \mu^{\delta^{13}C}_{t_n},\ \mu^{\log\text{CO}_2}_{t_n}\big)' \in \R^{3\times 1}.
\]
The measurement equation is scalar,
\[
y_n = c_{(n)} + \underbrace{Z_n}_{1\times 3}\, \mu_{t_n} + \eps_n,
\qquad Z_n = e_{p(n)}^{\top},
\qquad \eps_n \sim \mathsf{N}(0,\, h_n),
\]
where $e_p$ is the $p$-th standard basis vector of $\R^3$, so that $Z_n$ selects the observed proxy (for instance $Z_n = [\,0\ 1\ 0\,]$ at a $\delta^{13}$C observation). The intercept and variance are source-specific: $c_{(n)} = c^{p(n)}_{s(n)}$ and $h_n = \sigma^2_{\eps,s(n),p(n)}$ for the isotopes, and $c_{(n)} = c_{g(n)}$ and $h_n = \sigma^2_{\eps,g(n)}$ for CO$_2$, with $c = 0$ at the reference site and at the \texttt{b\_ca} reference group.

The transition equation propagates the full state between consecutive steps,
\[\arraycolsep=2pt\def\arraystretch{1.3}
\underbrace{\left[\begin{array}{c}
\mu^{\delta^{18}O}_{t_n+\Delta t_n}\\ \mu^{\delta^{13}C}_{t_n+\Delta t_n}\\ \mu^{\log\text{CO}_2}_{t_n+\Delta t_n}
\end{array}\right]}_{\mu_{t_n+\Delta t_n}}
=
\underbrace{\left[\begin{array}{ccc} 1&0&0\\ 0&1&0\\ 0&0&1 \end{array}\right]}_{T \,=\, I_3}
\mu_{t_n}
+
\underbrace{\left[\begin{array}{ccc}
b_{11,k} & b_{21,k} & b_{31,k}\\
b_{12,k} & b_{22,k} & b_{32,k}\\
b_{13,k} & b_{23,k} & b_{33,k}
\end{array}\right]}_{B_k \,\in\, \R^{3\times 3}}
\underbrace{\left[\begin{array}{c} \Delta e_{t_n}\\ \Delta\varepsilon_{t_n}\\ \Delta\bar\omega_{t_n} \end{array}\right]}_{\Delta o_n}
+\ \eta_{k,t_n},
\]
where $T = I_3$ is the identity (random walk), $\Delta o_n$ collects the increments of the La2004 orbital drivers over the interval (Section~\ref{S:TriBIC}), and the rows of the orbital matrix $B_k$ correspond to the three proxies and its columns to the three drivers (eccentricity, obliquity, precession), so that row $j$ reproduces line $j$ of equation~\eqref{E:TriTrans}. The matrix is climate-state specific; $B_k = 0$ for all $k$ recovers the baseline random walk and $B_k \equiv B$ the constant-coefficient specification. The state disturbance is Gaussian with the climate-state rate matrix scaled by the time increment,
\[\arraycolsep=2pt\def\arraystretch{1.5}
\eta_{k,t_n} \sim \mathsf{N}(0,\, Q_k\,\Delta t_n), \qquad
Q_k =
\left[\begin{array}{ccc}
\sigma^2_{\eta,1,k} & \rho_{12,k}\,\sigma_{\eta,1,k}\sigma_{\eta,2,k} & \rho_{13,k}\,\sigma_{\eta,1,k}\sigma_{\eta,3,k}\\
\rho_{12,k}\,\sigma_{\eta,1,k}\sigma_{\eta,2,k} & \sigma^2_{\eta,2,k} & \rho_{23,k}\,\sigma_{\eta,2,k}\sigma_{\eta,3,k}\\
\rho_{13,k}\,\sigma_{\eta,1,k}\sigma_{\eta,3,k} & \rho_{23,k}\,\sigma_{\eta,2,k}\sigma_{\eta,3,k} & \sigma^2_{\eta,3,k}
\end{array}\right],
\]
where the proxy indices $1,2,3$ again denote $\delta^{18}$O, $\delta^{13}$C, $\log\text{CO}_2$ and $Q_k$ is the rate matrix of equation~\eqref{E:Qmatrix}. When successive observations share a time stamp, $\Delta t_n = 0$ and no disturbance is added. The orbital term is deterministic and leaves the Kalman filter and smoother recursions otherwise unchanged.

\subsection{Trivariate IWN(2)}

The pure $m$-fold integrated random walk of order $m = 2$ (Section~\ref{S:TriIWN2}) augments each proxy with a slope component. Stacking level and slope for the three proxies gives the six-dimensional state
\[
\alpha_{t_n} = \big(\mu^{\delta^{18}O}_{t_n}, \nu^{\delta^{18}O}_{t_n},\ \mu^{\delta^{13}C}_{t_n}, \nu^{\delta^{13}C}_{t_n},\ \mu^{\log\text{CO}_2}_{t_n}, \nu^{\log\text{CO}_2}_{t_n}\big)' \in \R^{6\times 1}.
\]
Only the level of the observed proxy is measured, so the scalar measurement loading selects it,
\[
Z_n = e_{p(n)}^{\top} \otimes [\,1\ \ 0\,] \in \R^{1\times 6}
\qquad(\text{e.g. } Z_n = [\,0\ 0\ 1\ 0\ 0\ 0\,]\ \text{at a } \delta^{13}\text{C observation}),
\]
with the same intercept $c_{(n)}$ and measurement variance $h_n$ as in the random-walk case. The transition matrix is block diagonal across the three proxies, each block the $2\times 2$ companion form of the once-integrated random walk,
\[\arraycolsep=2pt\def\arraystretch{1.3}
T_n = I_3 \otimes \left[\begin{array}{cc} 1 & \Delta t_n \\ 0 & 1 \end{array}\right].
\]
A single white-noise innovation enters each proxy's slope; cross-proxy dependence is carried by a $3\times 3$ slope-innovation covariance of the same form as $Q_k$, with the per-period slope-innovation standard deviations fixed at their univariate values (Section~\ref{S:TriIWN2}). The exact discrete-time covariance of the $(\mu,\nu)$ pair over an interval $\Delta t_n$ is the standard order-$m=2$ integrated-random-walk form given in \cite{bhkl2024}. The Milankovitch term enters as before, adding the constant orbital contribution $B\,\Delta o_n$ to the three level rows of the transition.

\bibliographystyle{plainnat}
\bibliography{co2_trivariate}

\end{document}